\documentclass[usenatbib,usedcolumn,useAMS]{mn2e}
\usepackage{textcomp}
\usepackage{amsmath}
\usepackage{array}
\usepackage{booktabs}
\usepackage{graphicx}
\usepackage{url}
\usepackage{lscape}
\usepackage{longtable}
\usepackage{graphicx}   
\usepackage{amsmath}    
\usepackage{amssymb}    

\begin{document}
\newcommand{\kms}{km\ s$^{-1}$}
\newcommand{\kmsmpc}{km\ s$^{-1}$\ Mpc$^{-1}$}
\newcommand{\ergscm}{erg\ s$^{-1}$ cm$^{-2}$} 
\newcommand{\ergs}{erg\ s$^{-1}$} 
\newcommand{\ergcms}{erg\ s$^{-1}$ cm$^{-2}$} 
\newcommand{\OII}{[\ion{O}{2}]}
\newcommand{\Deg}{^{\circ}}
\newcommand{\DFK}{D$_{\rm n}$(4000)}
\newcommand{\Hd}{H$\delta$}

\newcommand\aj{AJ} 
\newcommand\araa{ARA\&A} 
\newcommand\apj{ApJ} 
\newcommand\apjl{ApJ} 
\newcommand\apjs{ApJS} 
\newcommand\ao{Appl.~Opt.} 
\newcommand\apss{Ap\&SS} 
\newcommand\aap{A\&A} 
\newcommand\aapr{A\&A~Rev.} 
\newcommand\aaps{A\&AS} 
\newcommand\nat{Nature} 
\newcommand\pasp{PASP} 
\newcommand\pasj{PASJ} 
\newcommand\procspie{Proc.~SPIE} 
\newcommand\memras{MmRAS} 
\newcommand\mnras{MNRAS} 

\def\plottwo#1#2{\centering \leavevmode
    \includegraphics[angle=0,width=0.98\columnwidth]{#1} \hfil
    \includegraphics[angle=0,width=0.98\columnwidth]{#2}}

\title{Evaluating Tests of Virialization and Substructure Using Galaxy Clusters in the ORELSE Survey}

\author[N. Rumbaugh et al.]{
N. Rumbaugh$^{1}$,
B. C. Lemaux$^{1}$,
A. R. Tomczak$^{1}$,
L. Shen$^{1}$,
D. Pelliccia$^{1}$,\newauthor
L. M. Lubin$^{1}$,
D. D. Kocevski$^{2}$,
P.-F. Wu$^{3}$,
R. R. Gal$^{4}$,
S. Mei$^{5,6,7}$,\newauthor
C. D. Fassnacht$^{1}$,
G. K. Squires$^{8}$\\
$^{1}$Department of Physics, University of California, Davis, 1 Shields Avenue, Davis CA 95616, USA\\
$^{2}$Department of Physics and Astronomy, Colby College, Waterville, ME 04901, USA\\
$^{3}$Max-Planck Institut f\"{u}r Astronomie, K\"{o}nigstuhl 17, D-69117 Heidelberg, Germany\\
$^{4}$University of Hawai'i, Institute for Astronomy, 2680 Woodlawn Drive, HI 96822, USA\\
$^{5}$LERMA, Observatoire de Paris, PSL Research University, CNRS, Sorbonne Universit\'es, UPMC Univ. Paris 06, F-75014 Paris, France\\
$^{6}$Universit\'{e} Paris Denis Diderot, Universit\'e Paris Sorbonne Cit\'e, 75205 Paris Cedex 13, France\\
$^{7}$Jet Propulsion Laboratory, Cahill Center for Astronomy \& Astrophysics, California Institute of Technology, 4800 Oak Grove Drive, Pasadena, California 91011, USA\\
$^{8}$Spitzer Science Center, California Institute of Technology, M/S 220-6, 1200 E. California Blvd., Pasadena, CA, 91125, USA
}

\maketitle

\begin{abstract}
We evaluated the effectiveness of different indicators of cluster virialization using 12 large-scale structures in the ORELSE survey spanning from $0.7<z<1.3$. We located diffuse X-ray emission from 16 galaxy clusters using {\it Chandra} observations. We studied the properties of these clusters and their members, using {\it Chandra} data in conjunction with optical and near-IR imaging and spectroscopy. We measured X-ray luminosities and gas temperatures of each cluster, as well as velocity dispersions of their member galaxies. We compared these results to scaling relations derived from virialized clusters, finding significant offsets of up to 3-4$\sigma$ for some clusters, which could indicate they are disturbed or still forming. We explored if other properties of the clusters correlated with these offsets by performing a set of tests of virialization and substructure on our sample, including Dressler-Schectman tests, power ratios, analyses of the velocity distributions of galaxy populations, and centroiding differences. For comparison to a wide range of studies, we used two sets of tests: ones that did and did not use spectral energy distribution fitting to obtain rest-frame colours, stellar masses, and photometric redshifts of galaxies. Our results indicated that the difference between the stellar mass or light mean-weighted center and the X-ray center, as well as the projected offset of the most-massive/brightest cluster galaxy from other cluster centroids had the strongest correlations with scaling relation offsets, implying they are the most robust indicators of cluster virialization and can be used for this purpose when X-ray data is insufficiently deep for reliable $L_X$ and $T_X$ measurements. 

\end{abstract}


\section{Introduction}
\label{sec:intro}

As the largest gravitationally bound structures in the universe, galaxy clusters are an important cosmological probe. For example, they can be used to test cosmological models by constraining parameters such as $\sigma_8$ or the dark energy equation of state via cluster abundances or their mass function \citep{vik09b, rozo10, allen11}. Their distribution is also an important observable for testing models of structure formation and evolution. 

It is often important in the study of galaxy clusters to know their mass. However, since most of their mass content is in the form of dark matter, indirect measures are necessary. While weak lensing offers highly reliable mass estimates in the presence of high resolution imaging, they become prohibitively expensive to obtain as redshift increases \citep[see, e.g.,][]{vonderLinden14}. More indirect proxies are often used, such as gas mass, X-ray luminosity or temperature. Scaling relations between these observables and the cluster mass are then used to estimate the latter parameter, but such relations have considerable scatter \citep[e.g.,][]{pratt09,ett12}, and are valid only under certain conditions, such as virialization and hydrodynamic equilibrium. 

To obtain accurate parameter estimates using cluster scaling relations, we need to understand how these relations apply to different parameters, their scatter, and to which clusters they can be appropriately applied. In this paper, we will focus on those relations that involve properties of the intra-cluster medium (ICM) gas. In the simplest case, the ICM is heated only gravitationally as it infalls into the cluster. This leads to simple ``self-similar" power-law relations between parameters, such as temperature and luminosity \citep{kaiser86}. However, studies of these relations have shown clusters tend to deviate from these naive relations, implying non-gravitational sources of heating, such as from active galactic nuclei (AGN) \citep{mark98,AE99,XW00,vik02}. Such AGN activity can result in substantial deviations from the canonical self-similar scaling relations (e.g., \citealt{hilton12}), deviations that can vary depending on the location of the AGN with respect to the ICM and the mass of the galaxy hosting the AGN activity \citep{stott12}. Further, such scaling relations assume clusters are relaxed. If a cluster is still in the process of forming, gravitational energy will still be in the process of converting to internal energy, or if it has recently been disturbed by a merger, it will have substantial additional non-gravitational sources of energy. Therefore, such clusters will likely deviate from the scaling relations. 

Because of these deviations, the use of scaling relations, such as for mass estimates, will yield inaccurate results for non-virialized clusters \citep[see, e.g.,][]{smith03,arn07,nagai07,mah08}. It is therefore important to identify these clusters to avoid bias in mass estimates from fitting to these relations. A number of methods have been used for this purpose, including the two-dimensional, projected X-ray emission distribution \citep[e.g.,][]{jel05,allen08,okabe10,mah13}, the Dressler-Shectman test of substructure \citep[e.g.,][]{DS88, hall04}, deviations from Gaussianity in member dynamics \citep{marceu10, fakebruno13, haines15}, and projected offsets between various peaks such as X-ray, Sunyaev-Zeldovich, and the brightest cluster galaxy (BCG) \citep{mann12, hashimoto14, rossetti16}. However, it is questionable how broadly applicable such methods are and under what circumstances, if any, they fail to identify non-virialized clusters.  For instance, it has been shown that certain types of cluster-cluster mergers can leave limited impact on the line-of-sight dynamics of member galaxies, which make them difficult to identify using the degree to which the member line-of-sight velocities depart from Gaussianity \citep{nate17}. Similarly, viewing angle effects can severely limit the ability of the Dressler-Shectman test and similar tests to detect significant sub-structure when it is present \citep{white10}. The lack of a comprehensive study to test the efficacy of these methods over a large range of cluster types and operational definitions limits their general application. 

In this paper, we seek to determine which methods of detecting non-virialized structures are most effective. With the Observations of Redshift Evolution in Large-Scale Environments \citep[ORELSE;][]{lubin09}, we have an extensive multiwavelength dataset with which to do so, including $\sim 50$\ spectroscopically confirmed member galaxies per cluster, on average. The ORELSE survey is a systematic search for large-scale structures (LSSs) around an original sample of 20 galaxy clusters in a redshift range of $0.6 < z < 1.3$, designed to study galaxy properties over a wide range of local and global environments. The survey currently consists of 16 LSSs. These structures consist of several superclusters (defined here as LSSs with three or more member clusters) and merging systems, while some of the initially targeted galaxy clusters are found to be relatively isolated systems. This sample provides a wide range of environments that house both virialized and non-virialized galaxy clusters. 

Twelve of the sixteen LSSs in the ORELSE survey have {\it Chandra} imaging of sufficient quality to study diffuse X-ray emission, each of which has been described in previous papers \citep[see][]{rumb12,rumb13,rumb17}. We combine the X-ray data of these 12 LSSs, which are succinctly summarized in Table \ref{strsumtab}, with our extensive optical and near-infrared (IR) imaging and spectroscopy, as well as spectral energy distribution (SED) fitting, to assemble 10 tests of virialization/substructure and to compare them to three scaling relations between galaxy cluster observables. For our cosmological model, we assume $\Omega_m=0.3$, $\Omega_{\Lambda}=0.7$, and $h = H_0/70 km s^{-1} Mpc^{-1}$. 

We first describe our observations and data reduction, including SED fitting, in Section \ref{sec:obsred}. In Section \ref{sec:CP}, we discuss the various galaxy cluster properties that we measured. This involves carrying out two separate, but parallel, analyses: one using the results of our SED fitting to measure higher order galaxy properties such as rest-frame colour and stellar mass, and one without. The purpose of this parallel analysis is for applicability to a wide range of studies, including those without the requisite imaging depth or panchromatic coverage necessary to perform SED fitting to derive photometric redshifts, rest-frame magnitudes, or stellar masses. We then discuss scaling relations between galaxy cluster observables in Section \ref{sec:SR}. In Section \ref{sec:ana}, we analyze the effectiveness of our virialization/substructure tests by measuring their correlations with offsets from the scaling relations. We discuss the results of these correlations and their implications for surveys using galaxy clusters scaling relations in Section \ref{sec:disc} as well as make suggestions for optimal observational strategies for future surveys. 

\begin{table*}
\caption{Properties of Observed ORELSE LSSs}
\label{strsumtab}
\begin{tabular}{llllrrrll}
\toprule
\footnotesize{LSS}
 & \footnotesize{$\langle z\rangle$}
 & \footnotesize{$z$ Lower}
 & \footnotesize{$z$ Upper}
 & \footnotesize{Num. of}
 & \footnotesize{Num. of}
 & \footnotesize{{\it Chandra}-}
 & \footnotesize{$\sigma$}
 & \footnotesize{Confirmed}  \\
   { }
 & { }
 & \footnotesize{Bound}
 & \footnotesize{Bound}
 & \footnotesize{Known}
 & \footnotesize{Known}
 & \footnotesize{detected}
 & \footnotesize{Range$^{\rm b}$}
 & \footnotesize{Members$^{\rm c}$} \\
   { }
 & { }
 & { }
 & { }
 & \footnotesize{Clusters$^{\rm a}$}
 & \footnotesize{Groups$^{\rm a}$}
 & \footnotesize{Clusters}
 & { }
 & { }\\
\midrule
SG0023 & 0.84 & 0.82 & 0.87 & 0 & 5 & 0 & 218-418 & 250 \\
RCS0224 & 0.77 & 0.76 & 0.79 & 2 & 0 & 1 &710-825 & 119 \\
SC0849 & 1.26 & 1.25 & 1.28 & 3 & 2 & 2 &260-840 & 111 \\
RXJ0910 & 1.11 & 1.08 & 1.15 & 2 & 0 & 2 & 724-840 & 142 \\
RXJ1053 & 1.14 & 1.10 & 1.15 & 1 & 0 & 0 & $898\pm142$ & 72 \\
RXJ1221 & 0.70 & 0.69 & 0.71 & 1 & 1 & 1 & 427-753 & 160 \\
SC1324 & 0.76 & 0.65 & 0.79 & 3 & 4 & 3 & 186-873 & 452 \\
Cl1350 & 0.80 & 0.79 & 0.81 & 1 & 2 & 1 & 300-802 & 102 \\
SC1604 & 0.90 & 0.84 & 0.96 & 5 & 3 & 3 & 287-772 & 531 \\
RXJ1716 & 0.81 & 0.80 & 0.83 & 3 & 0 & 1 & 624-1120 & 197 \\
RXJ1757 & 0.69 & 0.68 & 0.71 & 1 & 0 & 1 & $862\pm108$  & 74 \\
RXJ1821 & 0.82 & 0.80 & 0.84 & 1 & 0 & 1 & $1129\pm100$  & 131 \\
\bottomrule
\multicolumn{9}{p{13.5cm}}{$^{\rm a}$ \footnotesize{Clusters and groups are defined as having velocity dispersions greater than or less than 550 \kms, respectively.}}\\
\multicolumn{9}{p{13.5cm}}{$^{\rm b}$ \footnotesize{In units of \kms. For LSSs with more than group or cluster, this measurement is the range of velocity dispersions of groups and clusters within the LSS.  All velocity dispersions are measured within 1 $h_{70}^{-1}$ Mpc of the luminosity-weighted spectroscopic member center (see \citealt{ascaso14} for details).}}\\
\multicolumn{9}{l}{$^{\rm c}$ \footnotesize{Spectroscopically confirmed galaxies within the redshift bounds of the LSS.}}\\
\end{tabular}
\end{table*}

\section{Observations and Reduction}
\label{sec:obsred}

\subsection{Chandra Observations}

All X-ray imaging was conducted using the Advanced CCD Imaging Spectrometer (ACIS) of the {\it Chandra X-Ray Observatory}. Both the ACIS-I and ACIS-S arrays were used. These have fields of view of $16\farcm9\times16\farcm9$ and $8\farcm3\times50\farcm6$, respectively. In most cases, the ACIS-I field of view is extended through the inclusion of the closest 1-2 ACIS-S chips, and vice versa. Several of the larger structures, SC1324 and SC1604, were imaged with multiple pointings of the array. Characteristics of the individual observations are listed in Table \ref{obsIDtab}. While we planned observations to have exposure times of approximately 50 ks per pointing per field, we supplemented our sample with publicly available data. These observations were exposed for varying lengths and can be distinguished, in Table \ref{obsIDtab}, as those with a PI other than Lubin. Note that, though \emph{Chandra} observations for SG0023 and RXJ1053 are included in Table \ref{obsIDtab} for completeness, these fields are removed from our analysis due to insufficient counts necessary to reliably measure X-ray luminosity, temperature, or both for the known structures in these fields. 

The reduction of the data was conducted using the {\it Chandra} Interactive Analysis of Observation 4.7 software \citep[CIAO;][]{frusc06}. We used the Imperial reduction pipeline, which is described in detail in \citet{laird09} and \citet{nandra15}. For a summary of this reduction process, see \citet{rumb17}. 

\begin{table*}
\caption{{\it Chandra} Observations}
\label{obsIDtab}
\begin{tabular}{lllllll}
\toprule
\footnotesize{Observation}
 & \footnotesize{Target}
 & \footnotesize{Instrument}
 & \footnotesize{PI}
 & \footnotesize{Exposure}
 & \footnotesize{RA $^{\rm a}$}
 & \footnotesize{Dec. $^{\rm a}$} \\
   {ID}
 & { }
 & { }
 & { }
 & {Time (ks)}
 & { }
 & { }\\
\midrule
7914 & SG0023 & ACIS-I & Lubin & 49.38 & 00\ 23\ 52.30 & 04\ 22\ 34.20 \\
3181 & RCS0224 & ACIS-S & Gladders & 14.37 & 02\ 24\ 34.10 & -00\ 02\ 30.90 \\
4987 & RCS0224 & ACIS-S & Ellingson & 88.97 & 02\ 24\ 34.10 & -00\ 02\ 30.90 \\
927 & Cl0849 & ACIS-I & Stanford & 125.15 & 08\ 48\ 55.90 & 44\ 54\ 50.00 \\
1708 & Cl0849 & ACIS-I & Stanford & 61.47 & 08\ 48\ 55.90 & 44\ 54\ 50.00 \\
2227 & RXJ0910 & ACIS-I & Stanford & 105.74 & 09\ 10\ 45.41 & 54\ 22\ 05.00 \\
2452 & RXJ0910 & ACIS-I & Stanford & 65.31 & 09\ 10\ 45.41 & 54\ 22\ 05.00 \\
4936 & RXJ1053 & ACIS-S & Predehl & 92.4 & 10\ 53\ 43.00 & 57\ 35\ 00.00 \\
1662 & RXJ1221 & ACIS-I & van Speybroeck & 79.08 & 12\ 21\ 24.50 & 49\ 18\ 14.40 \\
9403 & SC1324 & ACIS-I & Lubin & 26.94 & 13\ 24\ 49.50 & 30\ 51\ 34.10 \\
9404 & SC1324 & ACIS-I & Lubin & 30.4 & 13\ 24\ 42.50 & 30\ 16\ 30.00 \\
9836 & SC1324 & ACIS-I & Lubin & 20 & 13\ 24\ 42.50 & 30\ 16\ 30.00 \\
9840 & SC1324 & ACIS-I & Lubin & 21.45 & 13\ 24\ 49.50 & 30\ 51\ 34.10 \\
2229 & Cl1350 & ACIS-I & Stanford & 58.31 & 13\ 50\ 46.10 & 60\ 07\ 09.00 \\
6932 & SC1604 & ACIS-I & Lubin & 49.48 & 16\ 04\ 19.50 & 43\ 10\ 31.00 \\
6933 & SC1604 & ACIS-I & Lubin & 26.69 & 16\ 04\ 12.00 & 43\ 22\ 35.40 \\
7343 & SC1604 & ACIS-I & Lubin & 19.41 & 16\ 04\ 12.00 & 43\ 22\ 35.40 \\
548 & RXJ1716 & ACIS-I & van Speybroeck & 51.73 & 17\ 16\ 52.30 & 67\ 08\ 31.20 \\
10443 & RXJ1757 & ACIS-I & Lubin & 21.75 & 17\ 57\ 19.80 & 66\ 31\ 39.00 \\
11999 & RXJ1757 & ACIS-I & Lubin & 24.7 & 17\ 57\ 19.80 & 66\ 31\ 39.00 \\
10444 & RXJ1821 & ACIS-I & Lubin & 22.24 & 18\ 21\ 38.10 & 68\ 27\ 52.00 \\
10924 & RXJ1821 & ACIS-I & Lubin & 27.31 & 18\ 21\ 38.10 & 68\ 27\ 52.00 \\
\bottomrule
\multicolumn{7}{l}{$^{\rm a}$ \footnotesize{Coordinates refer to those of the observation aim point.}}\\
\end{tabular}
\end{table*}

\subsection{Photometry and Spectroscopy}

We have photometric observations of our sample across a wide range of bands, from optical to near-infrared (NIR), typically $B/V/R/I/Z/J/K/[3.6]/[4.5]$, as well as 24$\mu$m data which we do not include for this study. This dataset is composed of both our own ORELSE observing campaigns and archival data. This includes data from the Large Format Camera \citep[LFC;][]{simcoe00} on the Palomar 200-inch Hale Telescope, Suprime-Cam on the Subaru Telescope \citep{Suprime-Cam}, MegaCam and the Wide-field InfraRed Camera \citep[WIRCam;][]{WIRCam} on the Canada France Hawaii Telescope (CHFT), the Wide Field Camera \citep[WFCAM;][]{WFCAM} on the United Kingdom InfraRed Telescope (UKIRT), and the InfraRed Array Camera \citep[IRAC;][]{IRAC} on the {\it Spitzer Space Telescope}. For full details on the reduction of these data see \citet{gal08} and \citet{adam17}. Additionally, SC1604 has imaging from the Advanced Camera for Surveys (ACS) on-board the Hubble Space Telescope (HST)\footnote{SG0023, RCS0224, SC0849, RXJ0910, RXJ1053, SC1324, RXJ1716, and RXJ1757 also have archival coverage with varying exposure times and spatial coverages from ACS and the Wide Field Planetary Camera 2 \citep[WFPC2;][]{WFPC2} which we do not include in this study.}. See \citet{koc09b} for the details of these ACS observations. 

Photometry in the optical through near-IR (B-K bands) are measured in fixed circular apertures on images that are convolved to the seeing of the image with the largest point spread function (PSF). The size of this circular aperture is set to be 1.3x the size of the largest PSF, a choice which tends to maximize the signal-to-noise ratio of the extracted flux \citep{whit11}. The image used to detect sources on which photometric measurements were to be performed varied from field to field but were generally comprised of at least one image whose central filter was redder than the $D_{n}(4000)$/Balmer break at the redshift of the LSS in that field. A summary of detection images used along with the worst seeing across all of our ground-based images for all ORELSE fields studied in this paper is presented in Table \ref{tab:det}. For Spitzer/IRAC photometry we use T-PHOT \citep{merlin15}, a software package specifically designed to extract photometry in crowded imaging. Note that all optical through NIR photometry are used in our SED-fitting procedures. The 80\% completeness depth for all images of the two fields which were studied here and not included in \citet{adam17} are given in Table \ref{tab:limits}. A full description of the procedure of measuring photometry, estimating image depths, and the 80\% completeness depths for images taken on the remainder of the ORELSE fields studied here can be found in \citet{adam17}.

Our photometric catalog is complemented by extensive spectroscopic data. The bulk of these data are from the Deep Imaging Multi-Object Spectrograph \citep[DEIMOS;][]{Faber03} on the Keck II 10m telescope. DEIMOS has a wide field of view ($16\farcm9 \times 5\farcm0$), high efficiency, and is able to position over 120 targets per slit mask, ideal for establishing an extensive spectroscopic catalog. We used the 1200 line mm$^{-1}$ grating, blazed at 7500 \AA, and 1$\arcsec$-wide slits for a pixel scale of 0.33 \AA\ pixel$^{-1}$ and a FWHM resolution of $\sim1.7\ $\AA. Central wavelengths varied for each LSS, in the range 7000-8700 \AA, and the approximate spectral coverage was $\sim \pm1300$\ \AA\ around these central wavelengths. For an overview of the DEIMOS data used in this study see \citet{lem12} and \citet{rumb17}.

In addition to our own DEIMOS data, SG0023, RXJ0910, SC1604 and RXJ1821 had some spectral redshifts obtained from the Low-Resolution Imaging Spectrometer (LRIS; \citealt{oke95}). See \citet{oke98, GalLub04,gioia04,tan08} for further details on these observations and their reduction. In the SC0849 supercluster we supplemented our DEIMOS observations with a large number of redshifts obtained by a variety of different telescopes and instruments. The bulk of the spectral redshifts of SC0849 member galaxies in the two clusters presented in this paper were obtained with LRIS \citep{stan97, ros99, mei06}. These data were complemented by redshifts that were obtained by observations with a combination of the Faint Object Camera and Spectrograph (FOCAS; \citealt{kash02}) on Subaru, the northern version of the Gemini Multi-Object Spectrographs (GMOS-N; \citealt{hook04}), and DEIMOS using the 600 l mm$^{-1}$ grating, though these observations were primarily aimed at the surrounding large scale structure. Depending on the analysis, 40-45\% of the secure spectral redshifts of the member population of the two SC0849 clusters presented in this paper are drawn from these observations, with our own redshifts obtained from DEIMOS making up the remainder. For more details on the observation and reduction of this supplementary SC0849 spectroscopic data see \citet{mei12} and references therein. For more details on the DEIMOS spectroscopy taken specifically for ORELSE, its reduction, and the process of measuring redshifts see \citet{lem17} and \citet{adam17} and references therein. Only secure spectroscopic redshifts were used in our analysis, meaning those with quality flags of $Q$ = -1, 3, 4 as defined in \citet{gal08} and \citet{new13}. To the best of our ability, this scheme was applied equally to our own DEIMOS data and other spectroscopic data incorporated into the analysis in this paper.

\begin{table}
    \caption{Photometric Information}
    \label{tab:det}
    \begin{center}
    \begin{tabular}{rccc}
    \hline
    Field & Detection & Detection  & Worst \\
          &   Image   & Instrument &  PSF  \\
    \hline
    RXJ0910 & $R_C,I_+,Z_+$ & Subaru/Suprime-Cam & 1.00\arcsec \\
    Cl1350 & $r$ & CFHT/MegaCam & 1.96\arcsec \\
    RXJ1757 & $r',i'$ & Palomar/LFC & 1.24\arcsec \\
    RXJ1821 & $Y$ & Subaru/Suprime-Cam & 1.23\arcsec \\
    RCS0224 & $I_+$ & Subaru/Suprime-Cam &  1.25\arcsec \\
    RXJ1053 & $Z_+$ & Subaru/Suprime-Cam & 1.30\arcsec \\
    RXJ1221 & $i'$ & Palomar/LFC & 1.37\arcsec \\
    RXJ1716 & $R_C,I_+,Z_+$ & Subaru/Suprime-Cam & 0.89\arcsec \\
    SC0849 & $Z_+$ & Subaru/Suprime-Cam & 1.40\arcsec \\
    SC1324 & $i$ & Palomar/LFC & 1.28\arcsec \\
    SC1604 & $R_C$ & Subaru/Suprime-Cam & 1.30\arcsec\\
    \hline
    \end{tabular}
    \end{center}
\end{table}

\begin{table*}
    \begin{center}
    \caption{Photometry}
    \label{tab:limits}
    {\vskip 1mm}
    \begin{tabular}{c @{\hskip 15mm} c}

        \begin{tabular}{llll}

        \hline \\[-3.3mm]

        Filter & Telescope & Instrument & Depth$^a$ \\[0mm]
	\hline \\[-3mm]
        Cl1350 \\[0.5mm]
        \hline \\[-3mm]
        \hline \\[-2mm]

        $B$    &   Subaru   &   Suprime-Cam   &   26.5   \\
        $V$    &   Subaru   &   Suprime-Cam   &   25.8   \\
        $R_C$  &   Subaru   &   Suprime-Cam   &   25.1   \\
        $g^{\prime}$    &   CFHT     &   MegaCam       &   24.4   \\
        $r^{\prime}$    &   CFHT     &   MegaCam       &   24.3   \\
        $r^{\prime}$   &   Palomar  &   LFC           &   25.0   \\
        $i^{\prime}$   &   Palomar  &   LFC           &   23.5   \\
        $z^{\prime}$   &   Palomar  &   LFC           &   22.9   \\
        $[3.6]$  &   {\it Spitzer}  &   IRAC   &   23.4   \\
        $[4.5]$  &   {\it Spitzer}  &   IRAC   &   23.4   \\[1mm]

        \hline \\[-3mm]

        RXJ1221 \\[0.5mm]
        \hline \\[-3mm]
        \hline \\[-2mm]

        $B$    &   Subaru   &   Suprime-Cam   &   26.6   \\
        $V$    &   Subaru   &   Suprime-Cam   &   26.1   \\
        $r'$   &   Palomar   &   LFC   &   24.2   \\
        $i'$   &   Palomar   &   LFC   &   24.4   \\
        $z'$   &   Palomar   &   LFC   &   22.8   \\
        $J$   &   UKIRT   &   WFCAM   &   22.4   \\
        $K$   &   UKIRT   &   WFCAM   &   21.9   \\
        $[3.6]$  &   {\it Spitzer}  &   IRAC   &   24.0   \\
        $[4.5]$  &   {\it Spitzer}  &   IRAC   &   23.8   \\[1mm]

	\end{tabular}

        &

        \begin{tabular}{llll}

	\hline \\[-3mm]

        Filter & Telescope & Instrument & Depth$^a$ \\[0mm]
		
	\hline \\[-3mm]
        RXJ1053 \\[0.5mm]
        \hline \\[-3mm]
        \hline \\[-2mm]

        $u^{\ast}$    &   CFHT     &   MegaCam       &   24.8   \\
        $g^{\prime}$    &   CFHT     &   MegaCam       &   25.7   \\
        $r^{\prime}$    &   CFHT     &   MegaCam       &   24.5   \\
        $z^{\prime}$    &   CFHT     &   MegaCam       &   23.6   \\
        $B$      &   Subaru   &   Suprime-Cam   &   26.1   \\
        $V$      &   Subaru   &   Suprime-Cam   &   26.1   \\
        $R_C$    &   Subaru   &   Suprime-Cam   &   25.2   \\
        $R_+$    &   Subaru   &   Suprime-Cam   &   26.4   \\
        $I_+$    &   Subaru   &   Suprime-Cam   &   25.1   \\
        $Z_+$    &   Subaru   &   Suprime-Cam   &   25.5   \\
        $J$   &   UKIRT   &   WFCAM   &   22.3   \\
        $K$   &   UKIRT   &   WFCAM   &   21.7   \\
        $[3.6]$  &   {\it Spitzer}  &   IRAC   &   23.9   \\
        $[4.5]$  &   {\it Spitzer}  &   IRAC   &   23.4   \\
        $[5.8]$  &   {\it Spitzer}  &   IRAC   &   21.7   \\
        $[8.0]$  &   {\it Spitzer}  &   IRAC   &   21.8   \\

        \end{tabular}

    \end{tabular}

    \end{center}

    $^a$ 80\% completeness limits derived from the recovery rate of artificial sources inserted at empty sky regions.

\end{table*}

\subsection{Spectral Energy Distribution Fitting}
\label{sec:SED}

We performed spectral energy distribution (SED) fitting for our sample using our optical to mid-infrared photometry, which we used to estimate the photometric redshifts of galaxies as well as properties such as stellar mass. To do this fitting, we used the Easy and Accurate $z_{phot}$ from Yale \citep[EAZY;][]{EAZY}, which performs an iterative $\chi^2$ fit using Projet d'Etude des GAlaxies par Synth\`{e}se \'{E}volutive \citep[P\'{E}GASE;][]{PEGASE} models, taking the results of our aperture photometry as input. This code outputs a probability density function $P\left(z\right)$, a measure of our confidence that the respective source is at a given redshift $z$. This PDF is modulated by a magnitude prior, designed to mimic the intrinsic redshift distribution for galaxies of given apparent magnitude. We adopted as the photometric redshift $z_{\rm{peak}}$, which is obtained by marginalizing over the output $P(z)$, except when an object has multiple significant peaks in its $P(z)$. In this case, the marginalization is constrained to the peak with the largest integrated probability.

To select a pure sample of galaxies, we cut sources that were likely stars, any objects with a S/N $<$ 3 in the detection band, those covered in less than five of the broadband images, ones that were saturated, or with catastrophic SED fits (defined as $\chi^2_{galaxy}>10$ from EAZY fits). To determine likely stars, another round of fitting with stellar templates was performed. See \citet{adam17} for more details.

To derive stellar masses and other galactic properties, we used the Fitting and Assessment of Synthetic Templates \citep[FAST;][]{FAST} code. FAST creates a multi-dimensional cube of model fluxes from a stellar population synthesis library (SPS). The best fit model is found by fitting each object to every object in this cube and minimizing $\chi^2$. High quality spectroscopic redshifts were used when available, and $z_{peak}$ from EAZY was used as a redshift prior for all other cases. See \citet{lem17} and \citet{adam17} for more details on the SED fitting. 

Besides deriving stellar masses, we also use the rest-frame colours derived from our SED fitting to separate galaxies into star-forming (SF) and quiescent populations. To divide the galaxies into these two categories, we use a two-colour selection technique proposed by \citet{ilbert10}. We adopted the rest-frame $M_{NUV} - M_{r}$ versus $M_{r} - M_{J}$ colour-colour diagram separations from \citet{lemaux14Herschel}. Galaxies at $0.5 < z \leq 1.0$ were considered quiescent if they had $M_{NUV} - M_r > 2.8(M_r - M_J) + 1.51$ and $M_{NUV} - M_r > 3.75$, while galaxies at $1.0 < z \leq 1.5$ were considered quiescent when $M_{NUV} - M_r > 2.8(M_r - M_J) + 1.36$ and $M_{NUV} - M_r > 3.6$. All other galaxies were classified as star-forming.

\section{Cluster Properties}
\label{sec:CP}

As discussed in Section \ref{sec:intro}, we seek to determine the most effective ways to determine the virialization status of galaxy clusters. To accomplish this, we first need to carry out a number of tests of virialization which we will then compare to offsets in the various empirical scaling relations drawn from the literature for our cluster sample presented in Section \ref{sec:SR}. In this section, we examine the properties of the individual clusters using optical and X-ray imaging and our spectroscopic data. We use these to carry out a number of tests of virialization and substructure on the galaxy clusters, and each cluster property discussed is used as part of one of these tests. 

Some of the following tests will use the results of the SED fitting, such as using the separation of galaxies into star-forming and quiescent populations to measure their separate velocity dispersions. This was made possible for this work trough extensive multiwavelength observations and time spent carrying out the fitting. Not all surveys have these resources and instead must rely on measures more closely related to observables, such as the observed optical colour or luminosity, as opposed to model-derived rest-frame colours or stellar mass. Where applicable, we carry out parallel tests: one using the results of our SED fitting, and one without those results.  

\begin{table*}
\caption{Cluster Properties}
\label{clusproptab}
\begin{tabular}{llllllll}
\toprule
\footnotesize{Cluster}
 & \footnotesize{$z$}
 & \footnotesize{Num. of}
 & \footnotesize{Gas Temp.}
 & \footnotesize{Bol. X-ray}
 & \footnotesize{Velocity}
 & \footnotesize{MMCG (BCG)}
 & \footnotesize{Quiescent} \\
   {}
 & {}
 & \footnotesize{Members$^{\rm a}$}
 & \footnotesize{(keV)}
 & \footnotesize{Lum. ($10^{44}$\ ergs/s)}
 & \footnotesize{Disp. (km/s)$^{\rm b}$}
 & \footnotesize{Vel. Off. (km/s)$^{\rm c}$}
 & \footnotesize{Fraction$^{\rm d}$}\\
\midrule
RCS0224B & 0.778 & 52 & $5.1^{+4.0}_{-1.7}$ & $2.0\pm0.1$ & $710\pm60$ & 1095 (1095) & $0.513\pm0.066$\\
SC0849C & 1.261 & 25 & $7.7^{+3.9}_{-2.2}$ & $2.4\pm0.2$ & $840\pm110$ & -196 (-196) & $0.427\pm0.067$\\
SC0849D & 1.270 & 23 & $4.0^{+6.3}_{-1.8}$ & $0.9\pm0.2$ & $700\pm110$ & 686 (686) & $0.292\pm0.067$\\
SC0910A & 1.103 & 23 & $2.9^{+1.7}_{-0.8}$ & $1.8\pm0.2$ & $840\pm240$ & -139 (-139) & $0.574\pm0.092$\\
SC0910B & 1.101 & 25 & $5.2^{+2.7}_{-1.4}$ & $2.4\pm0.2$ & $720\pm150$ & 757 (757) & $0.638\pm0.117$\\
RXJ1221B & 0.700 & 36 & $9.0^{+1.5}_{-1.1}$ & $11.0\pm0.3$ & $750\pm120$ & 3 (3) & $0.636\pm0.073$\\
SC1324A & 0.756 & 43 & $8.5^{+36.0}_{-4.3}$ & $2.0\pm0.2$ & $870\pm110$ & -413 (-413) & $0.458\pm0.088$\\
SC1324B & 0.698 & 13 & $^{\rm e}$ & $^{\rm e}$ & $680\pm140$ & 4 (4) & $0.630\pm0.117$\\
SC1324I & 0.696 & 27 & $2.9^{+4.1}_{-1.4}$ & $1.3\pm0.2$ & $890\pm130$ & -447 (-447) & $0.516\pm0.075$\\
Cl1350C & 0.800 & 43 & $4.7^{+1.7}_{-1.0}$ & $4.4\pm0.2$ & $800\pm80$ & -239 (-239) & $0.500\pm0.069$\\
SC1604A & 0.898 & 35 & $3.7^{+5.6}_{-1.8}$ & $2.4\pm0.3$ & $720\pm130$ & 126 (126) & $0.438\pm0.077$\\
SC1604B & 0.865 & 49 & $1.3^{+2.7}_{-0.5}$ & $1.6\pm0.2$ & $820\pm70$ & 33 (33) & $0.348\pm0.071$\\
SC1604D & 0.923 & 70 & $0.8^{+0.3}_{-0.4}$ & $1.3\pm0.9$ & $690\pm90$ & 50 (-199) & $0.410\pm0.082$\\
RXJ1716A & 0.809 & 83 & $6.4^{+1.2}_{-0.9}$ & $11.9\pm0.4$ & $1120\pm100$ & 2102 (2102) & $0.525\pm0.051$\\
RXJ1757 & 0.693 & 34 & $5.1^{+3.0}_{-1.6}$ & $2.0\pm0.2$ & $860\pm110$ & -960 (278) & $0.489\pm0.074$\\
RXJ1821 & 0.817 & 52 & $6.1^{+2.3}_{-1.4}$ & $7.5\pm0.4$ & $1120\pm100$ & 510 (510) & $0.512\pm0.049$\\
\bottomrule
\multicolumn{8}{l}{$^{\rm a}$ \footnotesize{Includes all spectroscopically confirmed members within a projected distance of 1 Mpc.}}\\
\multicolumn{8}{p{16cm}}{$^{\rm b}$ \footnotesize{Velocity dispersion measured using all spectroscopically confirmed cluster members within 1$h_{70}^{-1}$ Mpc of the X-ray centroid as the initial sample for $3\sigma$ clipping. See Section \ref{sec:veldisp} for more details.}}\\
\multicolumn{8}{l}{$^{\rm c}$ \footnotesize{MMCG (BCG) velocity offsets are measured relative to the velocity center of the galaxy cluster.}}\\
\multicolumn{8}{l}{$^{\rm d}$ \footnotesize{Quiescent fraction is corrected for selection effects. See Section \ref{sec:SF} for details.}}\\
\multicolumn{8}{p{16cm}}{$^{\rm e}$ \footnotesize{We were unable to measure an X-ray temperature for SC1324B with any meaningful precision. Since measuring X-ray luminosities involved using a model based on this X-ray temperature, we were unable to accurately measure a luminosity as well.}}\\
\end{tabular}
\end{table*}

\subsection{Weighted Mean Centers}
\label{sec:WMC}

One of the simplest ways of measuring the centroid for a galaxy cluster is to take the weighted mean of the positions of its members. A luminosity or stellar mass-weighted mean center can be useful as a test of virialization when compared to the X-ray centroid or position of the brightest/most massive cluster galaxy. The weighted mean centers are a measure of the position of galaxies within the cluster, so offsets with other centroids can be signs of a disturbed state. There are several ways to calculate these centers, though we only consider two here: weighting by the stellar mass and luminosity. For both measures, we included all galaxies spectroscopically confirmed in the redshift range $\langle z \rangle \pm 3\sigma_{v}$, where $\sigma_{v}$ is the galaxy line of sight differential velocity dispersion (see Section \ref{sec:veldisp}), and within $R_{proj}<1h_{70}^{-1}$ Mpc of the X-ray center\footnote{Note that some previous ORELSE studies, such as \citet{rumb13}, used red galaxy density peaks to center the LWMCs instead of the X-ray centers.}. We would expect that all cluster members would tend to be located within $1h_{70}^{-1}$ Mpc, and we investigate the impact of varying this radius in Section \ref{sec:norm}.

First, we use the galaxy stellar masses from our SED fitting described in Section \ref{sec:SED} to measure mass-weighted mean centers (MWMCs), which are given in Table \ref{centab}. The coordinates for these MWMCs for each cluster were calculated by:

\begin{multline}
\alpha_{MWMC} = \frac{\sum\limits_{i=1}^{N}\alpha_{i}\mathcal{M}_{\ast,i}}{\sum\limits_{i=1}^{N}\mathcal{M}_{\ast,i}}, \, \, \delta_{MWMC} = \frac{\sum\limits_{i=1}^{N}\delta_{i}\mathcal{M}_{\ast,i}}{\sum\limits_{i=1}^{N}\mathcal{M}_{\ast,i}}
\label{eqn:MWMC}
\end{multline}

\noindent where $\alpha_{MWMC, j}$ and $\delta_{MWMC, j}$ are the MWMC right ascension and declination for a given cluster and $N$ is the number of members of that cluster. However, stellar masses are not available in many studies. For comparison to such studies, we also measured luminosity-weighted mean centers (LWMCs). The luminosity weighting should be carried out ideally with a band with rest-frame coverage completely redward of $D_n(4000)$ as such coverage ensures that its flux density correlates well with stellar mass. We used the `supercolours', defined in \citet{rumb17}, which are parameterized rest-frame magnitude estimates using observed magnitudes and redshifts as inputs. They consist of $M_{red}$ and $M_{blue}$, which are created from the $r$, $i$, and $z$ band apparent magnitudes as an approximation of rest-frame magnitudes, and can be thought of as approximating $M_{u*}$ and $M_{B}$. The supercolours are easily derived from observed data according to the methodology of \citet{rumb17} without intensive model fitting. The supercolour fluxes were defined according to the redshift-dependent formulae:

\begin{multline}
f_{blue}=A_{blue}\left[1-B_{blue}\left(z-z_0\right)\right]f_{\nu,R}\\+\left(1-A_{blue}\right)\left[B_{blue}\left(z-z_0\right)\right]f_{\nu,I}\\
\label{eqn:supercolor1}
\end{multline}
\vspace{-0.2cm}
\begin{multline}
f_{red}=A_{red}\left[1-B_{red}\left(z-z_0\right)\right]f_{\nu,I}\\+\left(1-A_{red}\right)\left[B_{red}\left(z-z_0\right)\right]f_{\nu,Z}
\label{eqn:supercolor2}
\end{multline}

\noindent where $A_{blue}$, $B_{blue}$, $A_{red}$, $B_{red}$, and $z_0$ are the free parameters. This parameterization is set up such that $f_{blue}$ = $f_{nu,R}$ and $f_{red}$ = $f_{nu, I}$ at $z\sim0.65$ while $f_{blue} = f_{nu,I}$ and $f_{red} = f_{nu, Z}$ at $z\sim1.2$. For some parts of the analysis $f_{blue}$ and $f_{red}$ are transformed into their absolute magnitude equivalents, $M_{blue}$ and $M_{red}$, using the relations given in \citet{rumb17}. The exact values of the various constants in Equation \ref{eqn:supercolor2} were determined by fitting the parameterization to minimize the difference with the actual rest-frame colours on a subset of the data with SED fitting, and can be found in \citet{rumb17}. These two quantities can be thought of as reliable proxies of the rest-frame $u^{\ast}$$-$ and $B-$band flux densities of galaxies within the ORELSE redshift range of $0.55 < z < 1.4$ for cases where $riz$ imaging is available but full SED fitting is not possible. The accuracy with which these supercolours predict the true (SED-fit) rest-frame $u^{\ast}$$-$ and $B-$band flux densities are discussed in detail in \citet{rumb17}. In contrast to \citet{rumb17}, here we used the photometry that was measured using the methods of \citet{adam17} for all clusters to calculated $M_{blue}$ and $M_{red}$. As such, red-sequence fits in supercolor/magnitude space were re-measured using an identical approach to that of \citet{rumb17}. The quantity $f_{red}$ was used for the luminosity weighting for all clusters. The LWMC coordinates for a cluster were then calculated by:

\begin{multline}
\alpha_{LWMC} = \frac{\sum\limits_{i=1}^{N}\alpha_{i}f_{\rm{red}, i}}{\sum\limits_{i=1}^{N}f_{\rm{red}, i}}, \, \delta_{LWMC} = \frac{\sum\limits_{i=1}^{N}\delta_{i}f_{\rm{red}, i}}{\sum\limits_{i=1}^{N}f_{\rm{red}, i}}
\label{eqn:LWMC}
\end{multline}

\noindent with $\alpha$, $\delta$, and $N$ having the same meaning as in Equation \ref{eqn:MWMC}. These centroids are listed in Table \ref{centab}. 

\begin{table*}
\caption{Cluster Centroids}
\label{centab}
\begin{tabular}{lcccccccccc}
\toprule
\scriptsize{Cluster}
 & \scriptsize{X-ray}
 & \scriptsize{X-ray}
 & \scriptsize{LWM}
 & \scriptsize{LWM}
 & \scriptsize{MWM}
 & \scriptsize{MWM}
 & \scriptsize{BCG}
 & \scriptsize{BCG}
 & \scriptsize{MMCG}
 & \scriptsize{MMCG}\\
   {}
 & \scriptsize{R.A.}
 & \scriptsize{Dec.}
 & \scriptsize{R.A.}
 & \scriptsize{Dec.}
 & \scriptsize{R.A.}
 & \scriptsize{Dec.}
 & \scriptsize{R.A.}
 & \scriptsize{Dec.}
 & \scriptsize{R.A.}
 & \scriptsize{Dec.}\\
\midrule
\footnotesize{RCS0224B}  & \footnotesize{02:24:34} & \footnotesize{-00:02:26.6} & \footnotesize{02:24:34} & \footnotesize{-00:02:26.9} & \footnotesize{02:24:34} & \footnotesize{-00:02:25.0} & \footnotesize{02:24:34} & \footnotesize{-00:02:27.9} & \footnotesize{02:24:34} & \footnotesize{-00:02:27.9} \\
\footnotesize{SC0849C}  & \footnotesize{08:48:58} & \footnotesize{+44:51:56.2} & \footnotesize{08:48:59} & \footnotesize{+44:52:03.4} & \footnotesize{08:48:58} & \footnotesize{+44:52:00.3} & \footnotesize{08:48:59} & \footnotesize{+44:51:57.2} & \footnotesize{08:48:59} & \footnotesize{+44:51:57.2} \\
\footnotesize{SC0849D}  & \footnotesize{08:48:36} & \footnotesize{+44:53:46.6} & \footnotesize{08:48:37} & \footnotesize{+44:54:10.7} & \footnotesize{08:48:35} & \footnotesize{+44:53:37.4} & \footnotesize{08:48:36} & \footnotesize{+44:53:36.1} & \footnotesize{08:48:36} & \footnotesize{+44:53:36.1} \\
\footnotesize{SC0910A}  & \footnotesize{09:10:09} & \footnotesize{+54:18:57.0} & \footnotesize{09:10:06} & \footnotesize{+54:18:50.8} & \footnotesize{09:10:07} & \footnotesize{+54:18:53.7} & \footnotesize{09:10:09} & \footnotesize{+54:18:53.8} & \footnotesize{09:10:09} & \footnotesize{+54:18:53.8} \\
\footnotesize{SC0910B}  & \footnotesize{09:10:45} & \footnotesize{+54:22:07.4} & \footnotesize{09:10:44} & \footnotesize{+54:22:13.2} & \footnotesize{09:10:45} & \footnotesize{+54:22:19.7} & \footnotesize{09:10:46} & \footnotesize{+54:23:29.0} & \footnotesize{09:10:46} & \footnotesize{+54:23:29.0} \\
\footnotesize{RXJ1221B}  & \footnotesize{12:21:26} & \footnotesize{+49:18:30.7} & \footnotesize{12:21:26} & \footnotesize{+49:18:30.5} & \footnotesize{12:21:27} & \footnotesize{+49:18:22.5} & \footnotesize{12:21:29} & \footnotesize{+49:18:17.2} & \footnotesize{12:21:29} & \footnotesize{+49:18:17.2} \\
\footnotesize{SC1324A}  & \footnotesize{13:24:49} & \footnotesize{+30:11:27.9} & \footnotesize{13:24:49} & \footnotesize{+30:11:52.1} & \footnotesize{13:24:49} & \footnotesize{+30:11:53.1} & \footnotesize{13:24:49} & \footnotesize{+30:11:38.9} & \footnotesize{13:24:49} & \footnotesize{+30:11:38.9} \\
\footnotesize{SC1324B}  & \footnotesize{13:24:21} & \footnotesize{+30:12:31.3} & \footnotesize{13:24:21} & \footnotesize{+30:12:57.6} & \footnotesize{13:24:21} & \footnotesize{+30:12:57.0} & \footnotesize{13:24:21} & \footnotesize{+30:12:43.2} & \footnotesize{13:24:21} & \footnotesize{+30:12:43.2} \\
\footnotesize{SC1324I}  & \footnotesize{13:24:50} & \footnotesize{+30:58:28.7} & \footnotesize{13:24:49} & \footnotesize{+30:58:20.6} & \footnotesize{13:24:50} & \footnotesize{+30:58:26.3} & \footnotesize{13:24:49} & \footnotesize{+30:58:40.7} & \footnotesize{13:24:49} & \footnotesize{+30:58:40.7} \\
\footnotesize{Cl1350C}  & \footnotesize{13:50:48} & \footnotesize{+60:07:11.5} & \footnotesize{13:50:51} & \footnotesize{+60:06:56.0} & \footnotesize{13:50:51} & \footnotesize{+60:06:57.1} & \footnotesize{13:50:60} & \footnotesize{+60:06:08.5} & \footnotesize{13:50:60} & \footnotesize{+60:06:08.5} \\
\footnotesize{SC1604A}  & \footnotesize{16:04:24} & \footnotesize{+43:04:36.6} & \footnotesize{16:04:23} & \footnotesize{+43:04:55.4} & \footnotesize{16:04:22} & \footnotesize{+43:04:57.2} & \footnotesize{16:04:24} & \footnotesize{+43:04:37.5} & \footnotesize{16:04:24} & \footnotesize{+43:04:37.5} \\
\footnotesize{SC1604B}  & \footnotesize{16:04:26} & \footnotesize{+43:14:23.5} & \footnotesize{16:04:27} & \footnotesize{+43:14:24.8} & \footnotesize{16:04:26} & \footnotesize{+43:14:16.5} & \footnotesize{16:04:26} & \footnotesize{+43:14:18.8} & \footnotesize{16:04:26} & \footnotesize{+43:14:18.8} \\
\footnotesize{SC1604D}  & \footnotesize{16:04:34} & \footnotesize{+43:21:07.1} & \footnotesize{16:04:33} & \footnotesize{+43:21:03.0} & \footnotesize{16:04:33} & \footnotesize{+43:21:02.5} & \footnotesize{16:04:36} & \footnotesize{+43:21:57.2} & \footnotesize{16:04:35} & \footnotesize{+43:21:56.0} \\
\footnotesize{RXJ1716A}  & \footnotesize{17:16:49} & \footnotesize{+67:08:24.4} & \footnotesize{17:16:51} & \footnotesize{+67:08:38.1} & \footnotesize{17:16:50} & \footnotesize{+67:08:36.8} & \footnotesize{17:16:49} & \footnotesize{+67:08:21.6} & \footnotesize{17:16:49} & \footnotesize{+67:08:21.6} \\
\footnotesize{RXJ1757}  & \footnotesize{17:57:19} & \footnotesize{+66:31:27.8} & \footnotesize{17:57:20} & \footnotesize{+66:31:16.2} & \footnotesize{17:57:21} & \footnotesize{+66:31:01.6} & \footnotesize{17:57:20} & \footnotesize{+66:31:32.6} & \footnotesize{17:57:21} & \footnotesize{+66:29:44.7} \\
\footnotesize{RXJ1821}  & \footnotesize{18:21:32} & \footnotesize{+68:27:55.4} & \footnotesize{18:21:31} & \footnotesize{+68:28:03.5} & \footnotesize{18:21:31} & \footnotesize{+68:28:08.3} & \footnotesize{18:21:31} & \footnotesize{+68:29:28.8} & \footnotesize{18:21:31} & \footnotesize{+68:29:28.8} \\
\bottomrule
\multicolumn{11}{p{16cm}}{\scriptsize{The acronyms LWM, MWM, BCG, and MMCG stand for, respectively, luminosity-weighted mean, mass-weighted mean, brightest cluster galaxy, and most massive cluster galaxy.}}
\end{tabular}
\end{table*}

\subsection{Most Massive and Brightest Cluster Galaxies}
\label{sec:MMCG}

We found both the most massive cluster galaxy (MMCG) and brightest cluster galaxy (BCG) for each cluster in our sample. Examining these galaxies is useful, since a BCG or MMCG with large positional or velocity offset from other centroiding measures can be indicative of a recent cluster-cluster merger \citep{bird94,GB02}. For the MMCGs, we used the results of our SED fitting, while, for the BCGs, we used the observed luminosities of the galaxies as an alternative to stellar mass, appropriate for comparison with studies that do not have SED fitting available. 

\subsubsection{MMCGs}
\label{sec:MMCG2}

We selected as potential MMCGs both galaxies that had secure spectroscopic redshifts and those that did not, but had photometric redshifts derived from our SED fitting described in Section \ref{sec:SED}. We used as our full sample galaxies within a projected distance of $R_{proj}<1.5R_{vir}$ from the X-ray centroids\footnote{We use this expanded projected radius to ensure we locate the BCG/MMCG for each cluster, even if it has an exceptionally large offset from the cluster center. In practice, the MMCG/BCG is always $R_{proj}\la1R_{vir}$} in projection from the X-ray centroid and typically much less ($R_{proj}<0.25R_{vir}$).. We required spectroscopically confirmed candidates to be within the redshift range $\langle z \rangle \pm 3\sigma_{v}$, where $\sigma_{v}$ is the galaxy line-of-sight differential velocity dispersion (see Section \ref{sec:veldisp}).

As our spectroscopic coverage is not 100\% complete in these clusters, it was necessary to supplement the spectroscopic member sample with potential members that had no secure spectral redshift, but did have a photometric redshift consistent with the cluster redshift. The allowed redshift range for potential cluster members with only photometric redshifts was expanded to $z_{min}-\sigma_{\Delta z/(1+z)}\left(1+z_{min}\right)$ and $z_{max}+\sigma_{\Delta z/(1+z)}\left(1+z_{max}\right)$ to account for the relative lack of precision of $z_{phot}$ measurements, where $z_{min}$ and $z_{max}$ refer to the minimum and maximum redshifts of the spectroscopic member redshift range, respectively. Values of $\sigma_{\Delta z/(1+z)}$ were estimated on a field by field basis by fitting a Gaussian to the
distribution of $\left(z_{spec} - z_{phot}\right)/\left(1 + z_{spec}\right)$ measurements in the range $0.5 < z < 1.2$ for all galaxies with a secure spectroscopic redshift (for more details see \cite{adam17}. The average $\sigma_{\Delta z/(1+z)}$ for all fields is $\sim0.025$, meaning we allow, on average, photometric-redshift members to spread in velocity an additional $\sim\pm$3500 km s$^{-1}$ relative to the spectroscopic members at the mean redshift of our cluster sample. This  photometric redshift range is chosen to maximize the product of purity and completeness of member galaxies, as derived from tests using spectroscopically confirmed samples.

For all cluster members at $z \leq 0.96$ we selected as potential MMCGs only those objects with $18.5 \leq i^{\prime}/I_{c}/I^{+} \leq 24.5$, while at $z>0.96$, we used $z^{\prime}/Z^{+} \leq 24.5$ instead. Due to the incompleteness of our spectroscopy, we include only objects with stellar masses of $M \geq 10^{10} M_{\odot}$, as this is roughly the stellar mass limit where our spectroscopic sample is representative of the underlying photometric sample at these redshifts and subject to the magnitude cuts above (see \citep{lu17} for more details). Additionally, this limit is comparable to the stellar mass completeness limit of our imaging data for all galaxy types at all redshifts considered in this paper (see \citealt{adam17}). 

We then selected the remaining galaxies with the top three stellar masses from our SED fits as the potential MMCG candidates for each cluster. For each MMCG candidate we inspected the SED fit and rejected any candidates with obvious photometric issues or probable stars. The most massive of the remaining MMCG candidates, all of which are spectroscopically confirmed, was then adopted as the MMCG for each cluster.

\subsubsection{BCGs}
\label{sec:BCG}

To select BCGs, we used the supercolours described in Section \ref{sec:WMC} and defined in \citet{rumb17}, though, as mentioned in \S\ref{sec:WMC} we now adopted the photometry measured using the methods of \citet{adam17} as input for all supercolor calculations in order to be consistent with our SED-fitting results. The BCG was chosen as the spectroscopically-confirmed galaxy\footnote{While, in principle, we are setting up this analysis to be independent of results from the SED fitting, we verified that each BCG in the spectroscopic member sample was brighter than all photometric redshift members, i.e., that it was the true BCG.} with the smallest value of $M_{red}$. The photometry as measured on our ground-based Suprime-Cam or LFC imaging was used in all cases to compute $M_{red}$ and identify the BCG. While our spectroscopy is nearly complete, photometric redshift analysis or careful colour/magnitude selection may be necessary in other cases to locate potential BCGs for less complete surveys. Note that only two clusters had BCGs that were not identified as the MMCG. In these two clusters, the average ratio of the stellar masses between the MMCG and the BCG was 2.09 and the difference between the $M_{red}$ of the MMCG and the BCG was small ($<$0.05 mags). Since many of the clusters in our sample have yet to form a truly dominant galaxy, a small luminosity gap typically exists between the BCG and the next brightest galaxies (see \citealt{ascaso14} and \citealt{rumb17}). Further, since $M_{red}$ is an imperfect proxy of stellar mass, a lack of concordance between the identified BCGs and MMCGs is to be expected at some level. Regardless, given the large overlap between the two samples and the fact that both sets of galaxies lie are extremely massive at these redshifts and generally appear on the red sequence, it is likely that both sets of galaxies have had considerable time to interact with the cluster potential. As such, it is likely the spatial and velocity information of both the BCGs identified and not identified as the MMCGs provide some level of information on the virialization state of the cluster regardless of their precise identity. 

\subsection{Velocity Dispersions}
\label{sec:veldisp}

Examining the velocity information of galaxy cluster members is useful for tests of virialization and substructure. Comparing velocity centers and dispersions of subpopulations (e.g., red vs. blue galaxies) provides tests of clusters' dynamical state and substructure \citep[e.g.,][]{ZF93}. Before studying subpopulations, we first examine the cluster velocity distributions as wholes and describe our measurement methods. 

We measure differential line-of-sight galaxy velocity dispersions (hereafter referred to simply as velocity dispersions) following the methods described in \citet{lubin02}, \citet{gal05}, and \citet{rumb13}. Unlike the values reported in Table \ref{strsumtab}, we include here all galaxies within 1$h_{70}^{-1}$\ Mpc of the X-ray center of each detected cluster. Adopting a different centroid for the defining cluster members, e.g., the LWMC, does not significantly affect the calculated velocity dispersions. 

To measure the velocity dispersions, we first select an initial redshift range by eye based on the redshift histogram. We perform iterative $3\sigma$ clipping, using the biweight scale estimator or gapper as defined in \citet{beers90}. The velocity dispersion measurement is given by the final iteration, and uncertainties are estimated using jackknife confidence intervals. Our measurements are presented in Table \ref{clusproptab}. 

\begin{figure*}
    \includegraphics[trim={8cm 11cm 10cm 11cm},clip,width=0.9\textwidth]{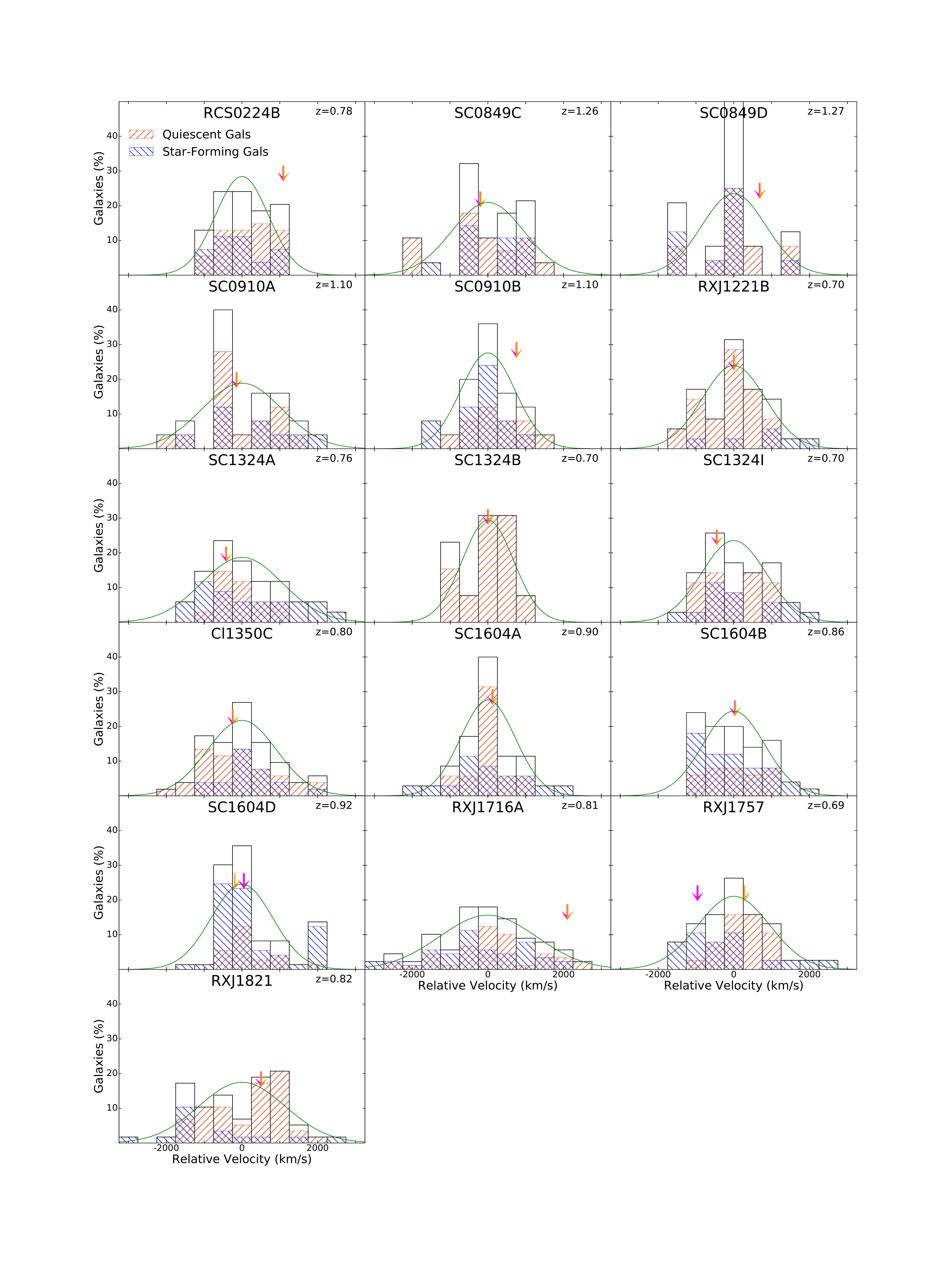}    
    \caption{
 Velocity histograms for each cluster are shown, relative to the cluster velocity center. Additionally, velocity histograms for the star-forming and quiescent subpopulations are shown in blue and red, respectively. To provide a sense of how well the velocity histogram conforms to a normal profile, a Gaussian distribution is overplotted with $\sigma=\sigma_v$. The velocities of the MMCG and BCG are shown with an arrow and half arrow, respectively.}
    \label{fig:veldisp}
\end{figure*}

Velocity histograms for each cluster are shown in Figure \ref{fig:veldisp}. The total distribution is shown, as well as those of the quiescent and star-forming populations, using hatched histograms. Velocities are given relative to the central redshift of the cluster defined as the mean redshift of all member galaxies. A Gaussian distribution with $\sigma$ equal to the velocity dispersion of the cluster and a mean value equal to that of the mean redshift of all spectral members is overplotted in each case. Note that these Gaussian distributions are shown for illustrative purposes and do not represent fits to the data. The velocities of the MMCGs and BCGs are shown with full and half arrows, respectively.

\subsubsection{Red vs. Blue and Quiescent vs. Star-Forming Galaxy Populations}
\label{sec:RB}

As stated in the previous subsection, studying the differences between the red and blue galaxy populations of a cluster can provide information on its substructure and dynamical state \citep[e.g.,][]{ZF93}. Roughly, the colour of the galaxies, in certain bands, can be seen as a proxy for the current star formation, since bluer galaxies (in rest-frame optical bands) tend to have younger stellar populations. A difference between the velocity dispersions of these populations can be a sign of virialization. Galaxies that have spent more time within a cluster have also had more time to feel the influence of dynamical friction, sending them closer to the core, as well as cluster processes such as ram pressure stripping, meaning they are more likely to have quenched star formation \citep{balogh01}. Galaxies which we know are star-forming are more likely to be infalling, residing closer to the cluster outskirts. Therefore, we would expect star-forming populations in virialized clusters to have larger velocity dispersions than the quiescent populations. Either way, this difference in velocity dispersions of subpopulations is supported by \citet{ZF93}, who find such differences between early and late-type galaxies, types which correlate well with quiescent and star-forming galaxies, respectively.

We can perform this test using our SED fitting results, using the cuts outlined in Section \ref{sec:SED} to classify galaxies as star-forming or quiescent. In addition, as part of our suite of virialization tests without SED fitting, we rely instead on some combination of the observed galaxy colours. For this latter case, we adopted the $M_{red}$ and $M_{blue}$ `supercolours' described in Section \ref{sec:WMC} for this purpose. To separate galaxies into red and blue populations, we performed red sequence fits on the member galaxies of the clusters of each LSS in $M_{red}$ versus $M_{red}-M_{blue}$ colour-magnitude space using the methods described in \citet{rumb17}. Red galaxies were defined as those above (i.e., redder than) the lower (bluer) edge of the red sequence, and blue galaxies were defined as everything below (bluer) than this edge. 

We calculated the velocity dispersions of the red, blue, star-forming, and quiescent populations using the method described in the previous subsection. The various subpopulation velocity dispersions are given in Table \ref{sigtab}, along with the number of galaxies in each population. Using these counts, we can compare the subdivisions of galaxies by observed colour and star formation status derived from SED fits. While the average percentage of star-forming galaxies is 45\%, the average percentage of blue galaxies is only 35\%, meaning some red galaxies are dusty enough to fall onto the red sequence on a colour-magnitude diagram, but are actually star-forming. Another possible explanation for this discrepancy is that our SED-fitting scheme failed for fainter, redder galaxies resulting in these galaxies being spuriously placed in the star-forming region. However, the median $M_{red}$ and $i$-band apparent magnitude of red star-forming galaxies is, in fact, brighter by 0.5 mags than the average blue star-forming galaxy, which broadly precludes this possibility. In addition, galaxies classified as red and star forming appear almost exclusively in the area of $NUVrJ$ phase space typically containing dustier galaxies (e.g., below the quiescent region but at $M_{r}-M_{J}>1$) and have stacked spectral properties which diverge from those of the overall star-forming population (lack of strong emission lines, presence of strong Balmer absorption) which are consistent with those of dusty star-forming galaxies (see, e.g., \citealt{lemaux14Herschel}). These two lines of evidence essentially rule out the possibility that the classification of such galaxies is spurious. While the star-forming versus quiescent results may more accurately categorize the cluster members, the difference is small enough that observed colour appears to be an acceptable proxy for star formation in the absence of SED fits, assuming the data and redshift range allow an analysis similar to our supercolours. In Figure \ref{fig:veldisp}, we plot the quiescent versus star-forming velocity histograms on top of the full histograms. 

In addition to examining the differences in velocity dispersions of subpopulations, we also examine the differences in velocity centers, which can be an indication of substructure \citep{ZF93}. We quantify the velocity centers by using the biweight location estimator defined in \citet{beers90} on the blue and red (or quiescent and star-forming) galaxies in each cluster. The estimates of the systemic velocities of each sets of sub-population were then differenced (red to blue and quiescent to star-forming). To estimate the significance of each of these velocity differences, we performed Monte Carlo simulations in which we randomly assigned the galaxies in each cluster to be red and blue (or quiescent and star-forming), while still preserving the true number of red and blue (or quiescent and star-forming) galaxies. The fraction of trials with a velocity difference larger than the observed difference is given in Table \ref{testtab} and serves as our estimate of its significance.

\begin{table*}
\caption{Velocity Dispersions}
\label{sigtab}
\begin{tabular}{lllllllllll}
\toprule
\footnotesize{Cluster}
 & \footnotesize{$\sigma_v^{\rm{a}}$}
 & \footnotesize{$N_{all}$}
 & \footnotesize{$\sigma_v$}
 & \footnotesize{$N_{Qui}$}
 & \footnotesize{$\sigma_v$}
 & \footnotesize{$N_{SF}$}
 & \footnotesize{$\sigma_v$}
 & \footnotesize{$N_{red}$}
 & \footnotesize{$\sigma_v$}
 & \footnotesize{$N_{blue}$}\\
   {}
 & \footnotesize{(All)}
 & {}
 & \footnotesize{(Qui.)}
 & {}
 & \footnotesize{(SF)}
 & {}
 & \footnotesize{(Red)}
 & {}
 & \footnotesize{(Blue)}
 & {}\\
   {}
 & \footnotesize{(km/s)}
 & {}
 & \footnotesize{(km/s)}
 & {}
 & \footnotesize{(km/s)}
 & {}
 & \footnotesize{(km/s)}
 & {}
 & \footnotesize{(km/s)}
 & {}\\
\midrule
RCS0224B & $700\pm60$ & 54 & $670\pm70$ & 32 & $740\pm150$ & 22 & $710\pm50$ & 42 & $540\pm240$ & 12 \\
SC0849C & $950\pm140$ & 28 & $1070\pm200$ & 17 & $910\pm140$ & 11 & $930\pm150$ & 19 & $890\pm320$ & 9 \\
SC0849D & $850\pm310$ & 24 & $810\pm250$ & 13 & $790\pm590$ & 11 & $670\pm120$ & 18 & $600\pm240$ & 6 \\
SC0910A & $1060\pm140$ & 25 & $980\pm150$ & 16 & $1110\pm280$ & 9 & $1020\pm160$ & 19 & $1150\pm430$ & 6 \\
SC0910B & $720\pm170$ & 25 & $700\pm160$ & 11 & $570\pm310$ & 14 & $830\pm200$ & 12 & $390\pm160$ & 12 \\
RXJ1221B & $830\pm90$ & 35 & $730\pm90$ & 29 & $980\pm600$ & 6 & $720\pm90$ & 28 & $1230\pm200$ & 7 \\
SC1324A & $1070\pm130$ & 34 & $680\pm160$ & 14 & $1270\pm150$ & 20 & $810\pm150$ & 19 & $1310\pm210$ & 15 \\
SC1324B & $680\pm140$ & 13 & $660\pm180$ & 12 & $0\pm0$ & 1 & $690\pm140$ & 12 & $0\pm0$ & 0 \\
SC1324I & $850\pm100$ & 35 & $690\pm80$ & 21 & $1120\pm300$ & 14 & $700\pm70$ & 26 & $1220\pm250$ & 9 \\
Cl1350C & $920\pm90$ & 52 & $1030\pm120$ & 34 & $650\pm140$ & 18 & $980\pm120$ & 38 & $720\pm160$ & 14 \\
SC1604A & $720\pm130$ & 35 & $460\pm110$ & 19 & $1070\pm230$ & 16 & $500\pm80$ & 23 & $1190\pm260$ & 12 \\
SC1604B & $820\pm80$ & 50 & $690\pm100$ & 18 & $890\pm110$ & 32 & $670\pm100$ & 26 & $740\pm160$ & 23 \\
SC1604D & $810\pm130$ & 73 & $410\pm180$ & 19 & $940\pm140$ & 54 & $510\pm110$ & 25 & $950\pm230$ & 45 \\
RXJ1716A & $1280\pm110$ & 92 & $1250\pm180$ & 44 & $1260\pm130$ & 45 & $1210\pm140$ & 57 & $1380\pm150$ & 35 \\
RXJ1757 & $950\pm130$ & 38 & $520\pm140$ & 20 & $1160\pm330$ & 18 & $690\pm110$ & 26 & $1430\pm290$ & 12 \\
RXJ1821 & $1140\pm90$ & 58 & $940\pm90$ & 44 & $1280\pm860$ & 14 & $920\pm80$ & 45 & $380\pm110$ & 9 \\
\bottomrule
\multicolumn{11}{p{15.3cm}}{$^{\rm a}$ \footnotesize{Velocity dispersion measured using all spectroscopically confirmed cluster members within 1$h_{70}^{-1}$ Mpc of the X-ray centroid as the initial sample for $3\sigma$ clipping. See Section \ref{sec:veldisp} for more details.}}\\
\end{tabular}
\end{table*}

\begin{table*}
\caption{Virialization and Substructure Tests}
\label{testtab}
\begin{tabular}{lllllllllllllll}
\toprule
\footnotesize{Cluster}
 & \footnotesize{$P\left(\Delta_v\right)$}
 & \footnotesize{$P\left(\Delta_v\right)$}
 & \footnotesize{$\Delta^{\rm b}$}
 & \footnotesize{$P\left(\Delta\right)^{\rm b}$}
 & \footnotesize{$P_3/P_0$}
 & \footnotesize{$P_4/P_0$}
 & \footnotesize{$P_3/P_0$}
 & \footnotesize{$P_3/P_0$}
 & \footnotesize{$P_4/P_0$}
 & \footnotesize{$P_4/P_0$}\\
   {}
 & \footnotesize{Red vs.}
 & \footnotesize{Qui. vs.}
 & {}
 & {}
 & {}
 & {}
 & \footnotesize{Upper}
 & \footnotesize{Lower}
 & \footnotesize{Upper}
 & \footnotesize{Lower}\\
   {}
 & {Blue$^{\rm a}$}
 & \footnotesize{SF$^{\rm a}$}
 & {}
 & {}
 & {}
 & {}
 & \footnotesize{Bound$^{\rm c}$}
 & \footnotesize{Bound$^{\rm c}$}
 & \footnotesize{Bound$^{\rm c}$}
 & \footnotesize{Bound$^{\rm c}$}\\
\midrule
RCS0224B & 0.953 & 0.106 & 57.830 & 0.001 & 0.189  & 0.139  & 5.158  & 0.611  & 2.943  & 0.169 \\
SC0849C & 0.373 & 0.721 & 43.437 & 0.129 & 3.182  & 0.009  & 10.023  & 1.415  & 2.470  & 0.177 \\
SC0849D & 0.839 & 0.572 & 17.659 & 0.091 & 6.773  & 11.743  & 23.825  & 3.789  & 21.305  & 5.067 \\
SC0910A & 0.204 & 0.544 & 12.768 & 0.060 & 1.714  & 0.469  & 11.695  & 0.832  & 3.122  & 0.521 \\
SC0910B & 0.505 & 0.301 & 13.648 & 0.000 & 0.504  & 0.092  & 7.598  & 1.081  & 3.054  & 0.278 \\
RXJ1221B & 0.354 & 0.017 & 37.456 & 0.774 & 1.882  & 0.974  & 5.247  & 1.074  & 2.846  & 0.414 \\
SC1324A & 0.779 & 0.535 & 47.083 & 0.139 & 28.410  & 1.432  & 54.699  & 10.962  & 13.057  & 1.169 \\
SC1324B & 0.779 & 0.000 & 6.179 & 0.185 & 5.978  & 4.223  & 37.457  & 3.622  & 17.817  & 1.148 \\
SC1324I & 0.157 & 0.818 & 40.650 & 0.362 & 3.601  & 1.686  & 24.365  & 2.514  & 14.421  & 1.142 \\
Cl1350C & 0.754 & 0.262 & 57.507 & 0.782 & 26.714  & 8.116  & 51.883  & 18.655  & 11.806  & 2.797 \\
SC1604A & 0.101 & 0.747 & 35.527 & 0.747 & 21.876  & 1.900  & 43.635  & 10.979  & 8.326  & 0.900 \\
SC1604B & 0.022 & 0.659 & 60.944 & 0.035 & 26.902  & 6.537  & 52.249  & 10.841  & 22.747  & 2.856 \\
SC1604D & 0.973 & 0.742 & 133.068 & 0.406 & 3.806  & 1.955  & 29.283  & 2.997  & 14.747  & 0.887 \\
RXJ1716A & 0.655 & 0.000 & 91.995 & 0.066 & 1.053  & 1.704  & 4.078  & 0.380  & 3.099  & 0.609 \\
RXJ1757 & 0.429 & 0.004 & 38.921 & 0.043 & 6.760  & 1.751  & 24.982  & 3.847  & 7.886  & 1.301 \\
RXJ1821 & 0.000 & 0.000 & 63.362 & 0.146 & 3.396  & 1.325  & 11.740  & 1.188  & 4.945  & 0.551 \\
\bottomrule
\multicolumn{11}{p{15.3cm}}{$^{\rm a}$ \footnotesize{Probabilities that the differences in velocity centers between the red/quiescent and blue/star-forming galaxy populations arose by chance.}}\\
\multicolumn{11}{p{15.3cm}}{$^{\rm b}$ \footnotesize{$\Delta$ is derived from the DS tests, and $P\left(\Delta\right)$ represents the likelihood that the null hypothesis of zero substructure is true (see Section \ref{sec:DStest}).}}\\
\multicolumn{11}{p{15.3cm}}{$^{\rm c}$ \footnotesize{Lower and upper bounds on $P_3/P_0$ and $P_4/P_0$ are the inner 68\% range of values of these power ratio parameters found in Monte Carlo simulations of Poisson noise based on each cluster's diffuse emission. See Section \ref{sec:DE} for more details.}}
\end{tabular}
\end{table*}

\subsection{Star-Forming and Quiescent Galaxy Fractions}
\label{sec:SF}

We may expect more virialized galaxy clusters to have more quiescent galaxy populations. As mentioned in Section \ref{sec:RB}, in a virialized cluster, galaxies tend to have spent more time close to the cluster core, where they are subjected to processes such as ram pressure stripping that can quench star formation. To look for a correlation between quiescence and virialization, as well as our other metrics, we measure the fraction of quiescent galaxies in each cluster, using the results of our SED fitting (see Section \ref{sec:SED}). 

We calculated the fraction of quiescent galaxies, $f_{q,comb}$, for each cluster using the sample of all galaxies with spectroscopic and photometric redshifts, to mitigate any observational bias associated with the former. As discussed in Section \ref{sec:MMCG}, galaxies with only photometric redshifts were considered as galaxy members when they were within $\sigma_{\Delta z/(1+z)}\left(1+z_{max/min}\right)$ of the spectroscopic member redshift range. Uncertainties in $f_{q,comb}$ were derived from Poissonian statistics. $f_q$ is given in Table \ref{clusproptab}. Note that calculation of $f_q$ is only possible with SED fitting, due to contamination of the red sequence by dusty star-forming galaxies and, more importantly, in the absence of photometric redshifts, the fraction of red galaxies is a complex function of the spectroscopic sampling rate, redshift success, and targeting strategy.

\subsection{Dressler-Shectman Tests of Substructure}
\label{sec:DStest}

As another test of substructure within a galaxy cluster, we use the Dressler-Shectman (D-S) test, defined in \citet{DS88} and \citet{hall04}. The D-S test estimates the degree of substructure present in a cluster using both spatial and velocity information, which has has the potential of being predictive in terms of the dynamical state of the cluster. This spatial and velocity information is incorporated into the statistic $\delta^2$, which is calculated for each individual galaxy based on its ten nearest neighbours:
\begin{equation}
\delta^2=\frac{11}{\sigma_v^2}\left[\left(v_{loc}-\bar{v}\right)^2+\left(\sigma_{loc}-\sigma_v\right)^2\right]
\end{equation}
\label{eq:DS}
where $\bar{v}$ and $\sigma_v$ are the mean velocity and velocity dispersion of the cluster as a whole. When calculating $\delta^2$ for a certain galaxy, $v_{loc}$ and $\sigma_{loc}$ are the mean velocity and velocity dispersion, respectively, of that galaxy and its ten nearest neighbours (which is the reason for the coefficient of 11 as the first term is set by N$_{nn}$+1). Choices other than ten neighbours have been used in computing the D-S statistic, especially in cases where the parent sample is not much larger than 10 \citep[see, e.g.,][]{pink96}. However, the choice of 10 here is appropriate as it approximately corresponds to the optimal window size for the typical number of member galaxies in the clusters in our sample \citep[see][]{silverman86}. To measure the level of substructure of a cluster as a whole, \citet{DS88} define $\Delta$, which is the sum of the $\delta$ statistics of each member galaxy. $\Delta$ has a distribution dependent on the specific set of coordinates and redshifts for the given sample, similar to $\chi^2$, with the expected value on the order of the number of cluster members. 

The results of our D-S tests are given in Table \ref{testtab}. To estimate the significance of the $\Delta$ values, we performed Monte Carlo simulations using the method of \citet{hall04} and \citet{rumb13}. For each trial, we shuffled the velocities, but not the spatial coordinates, of all cluster members, and recalculated $\Delta$. The fraction of trials with $\Delta$ larger than the observed value are also given in Table \ref{testtab}, as $P\left(\Delta\right)$. This value is our estimate of the likelihood that no substructure exists in the given cluster. We note that this likelihood only speaks to the possibility of substructure and not the virialization state of a given cluster. The relationship between the results of the D-S test to the virialization state will be investigated in Section \ref{sec:ana}. 

\subsection{Diffuse X-Ray Emission}
\label{sec:DE}

Examining the X-ray emission from a cluster provides information on the diffuse gas located at its center. A disturbed cluster can have asymmetries in its gas distribution or an offset between the center of the ICM emission and other centroiding measures.

To locate diffuse X-ray emission from the galaxy clusters in our sample, we first removed bright point sources from the {\it Chandra} images. The areas containing point source emission were filled in using Poisson distributions to simulate the background, with the background level estimated using an annulus around each object. We then convolved the {\it Chandra} images with azimuthally symmetric beta models of the form
\begin{equation}
f(r) = A\left(1+\frac{r^2}{r_c^2}\right)^{-3\beta+1/2}
\label{eq:betamodel}
\end{equation}

We used $\beta=2/3$ and core radii of $r_c=180$ kpc, which are typical for galaxy clusters \citep[see, e.g.,][]{AE99,ett04,maug06,hicks08}. We defined the centroids as the points of local maxima in the smoothed images. In Table \ref{centab}, we provide the coordinates of these X-ray centers\footnote{Any galaxy group or cluster not listed had diffuse emission which was not measurable at a high enough signal-to-noise ratio to be useful for our full analysis.}. 

\begin{figure*}
    \includegraphics[width=0.99\textwidth]{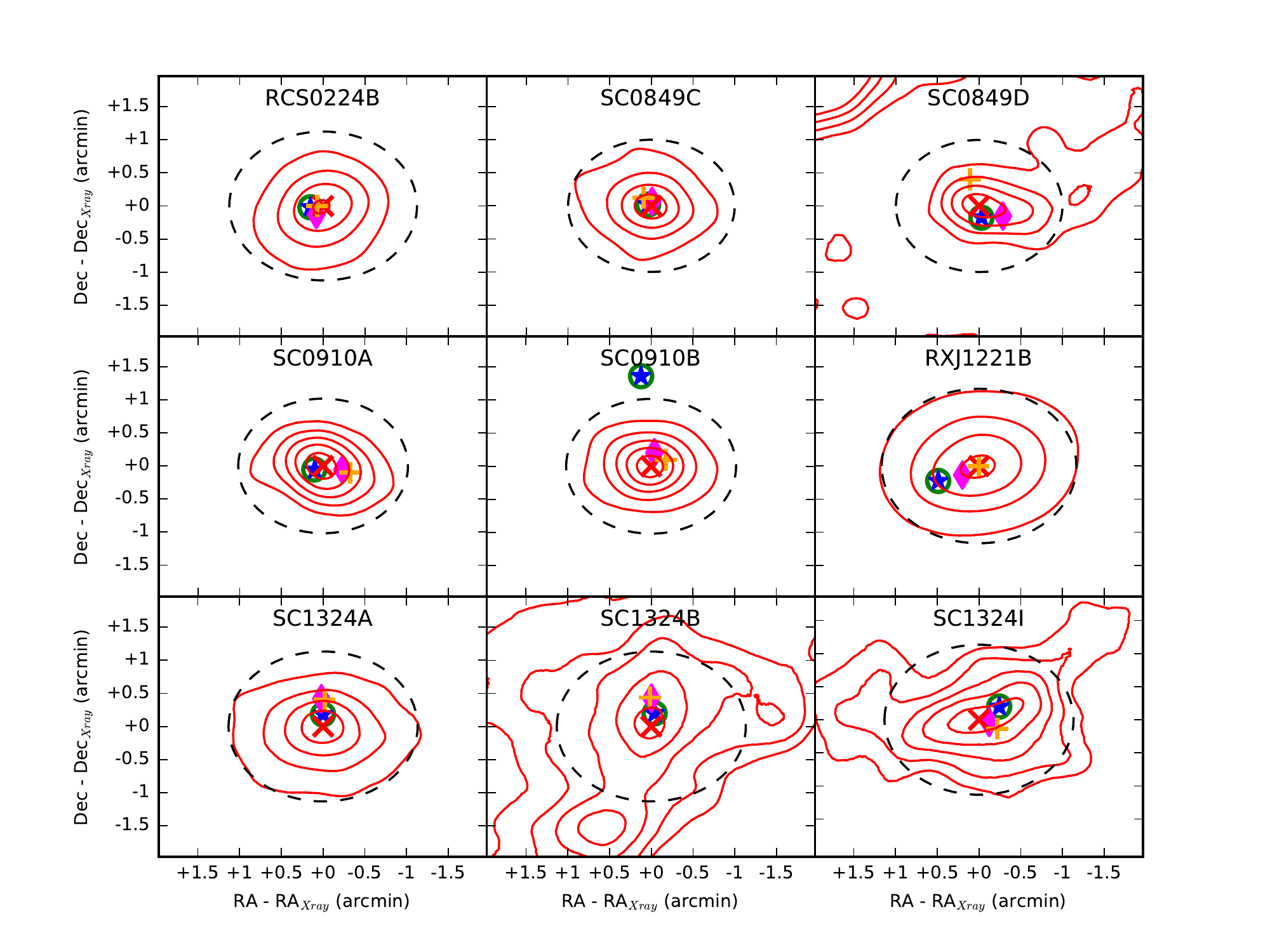}    
    \caption{
Spatial plots of diffuse X-ray emission are shown for each cluster, with different centroids marked. The center of the X-ray emission is marked with an X, the BCG is marked with an open circle, the MMCG is marked with a star, the luminosity-weighted mean center is marked with a plus, and the mass-weighted mean center is marked with a diamond. The dotted line is a circle of radius 0.5 Mpc at the redshift of the cluster, centered on the X-ray emission. The contours are derived from {\it Chandra} images of diffuse emission, convolved with a beta model. The background level was subtracted, and the image was divided by the RMS variability (see Section \ref{sec:DE} for more details). In these units, the contour levels are as follows: RCS0224B - 6, 12, 18, 24; SC0849C - 4, 8, 12, 16; SC0849D - 1.5, 3, 4.5, 6; SC0910A - 5, 10, 15, 20, 25; SC0910B - 2.5, 5, 7.5, 10, 12.5; RXJ1221B - 20, 40, 60, 80; SC1324A - 4, 8, 12, 16, 19; SC1324B - 2.5, 5, 7.5, 10, 12.5; SC1324I - 1, 2, 3, 4, 5. Contour levels were broadly set at 4-5 equally spaced intervals between the background and the peak of the X-ray emission.} 
    \label{fig:DE}
\end{figure*}

\begin{figure*}
    \includegraphics[width=0.99\textwidth]{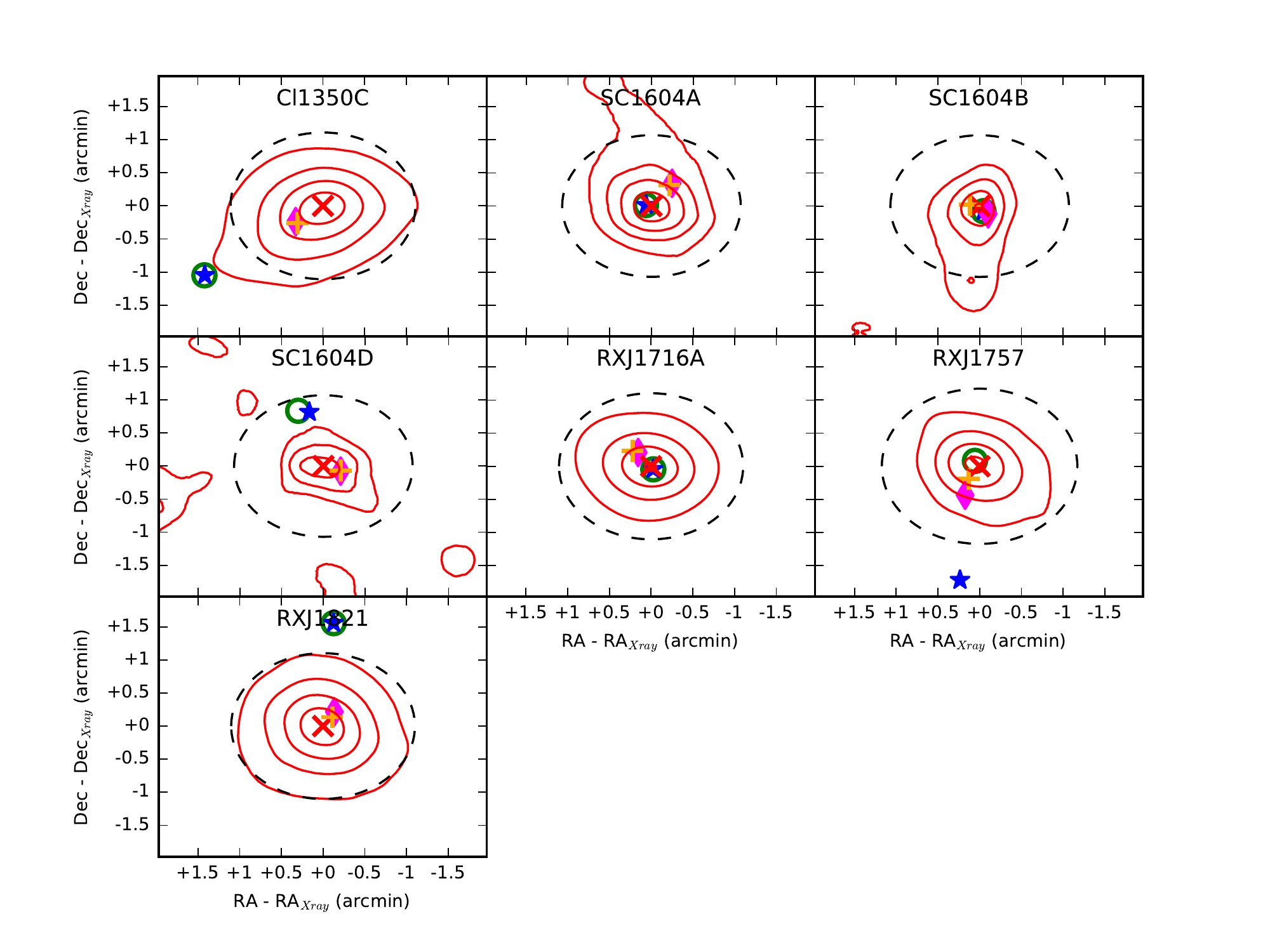}    
    \caption{
Spatial plots of diffuse X-ray emission are shown for each cluster, with different centroids marked. The center of the X-ray emission is marked with an X, the BCG is marked with an open circle, the MMCG is marked with a star, the luminosity-weighted mean center is marked with a plus, and the mass-weighted mean center is marked with a diamond. The dotted line is a circle of radius 0.5 Mpc at the redshift of the cluster, centered on the X-ray emission. The contours are derived from {\it Chandra} images of diffuse emission, convolved with a beta model. The background level was subtracted, and the image was divided by the RMS variability (see Section \ref{sec:DE} for more details). In these units, the contour levels are as follows: Cl1350C - 10, 20, 30, 40; SC1604A - 3, 6, 9, 12; SC1604B - 2, 4, 6, 7.5; SC1604D - 1, 2, 3; RXJ1716A - 30, 60, 90, 120; RXJ1757 - 7, 14, 21, 28; RXJ1821 - 10, 20, 30, 40.}
    \label{fig:DE2}
\end{figure*}

\subsubsection{Spatial Profiles}
\label{sec:spatprof}

X-ray contours from the smooth {\it Chandra} images are shown in Figures \ref{fig:DE} and \ref{fig:DE2}. While the contours appear largely symmetric, this can be quantified using power ratios, as in \citet{BT95,BT96}. The power ratios are basically calculated by evaluating the multipole moments of the surface brightness of the diffuse X-ray emission in a region around the cluster center. They are defined according to:
\begin{align}
P_0&=\left[a_0\ ln\left(R_{ap}\right)\right]^2\\
P_m&=\left(a_m^2+b_m^2\right)/\left(2m^2R_{ap}^{2m}\right)\\
a_m\left(R\right)&=\int_{r<R}\Sigma\left({\bf x}\right)r^m {\rm cos}\ m\phi\ d^2{\bf x}\\
b_m\left(R\right)&=\int_{r<R}\Sigma\left({\bf x}\right)r^m {\rm sin}\ m\phi\ d^2{\bf x}
\label{eq:PR}
\end{align}
where ${\bf x}=\left(r,\phi\right)$, $\Sigma\left({\bf x}\right)$ is the two-dimensional projection of mass density (in our case, we use photon counts per pixel as a proxy), and $R_{ap}$ is a chosen aperture radius inside which the power ratios are evaluated. Note that a perfectly round cluster will have $P_{m>1}$=0, so larger values will generally indicate more asymmetry. We measure the power ratios $P_3/P_0$ and $P_4/P_0$ for each cluster on the unsmoothed images, using $R_{ap} = 1 h_{70}^{-1}$ Mpc. These quantities for each cluster are given in Table \ref{testtab}. These two ratios should be most sensitive to asymmetries in the diffuse emission, and provide different information on the cluster gas, since odd moments are not sensitive to ellipticity \citep{jel05,don16}.  

To estimate uncertainties for the power ratios, we performed Monte Carlo simulations. {\it Chandra} images are very sparse, with either 0 or 1 photon counts in the overwhelming majority of pixels, so we first smoothed the {\it Chandra} images using a tophat kernel with a 5 pixel radius to get an estimate of the background level. We then generated a series of random images using Poisson functions based on the background level at each pixel\footnote{Uncertainties derived from an image with random noise placed on top of a sparsely sampled beta profile yielded similar values.}. The upper and lower bounds of the central 68\% of the Monte Carlo trials are shown in Table \ref{testtab}. Most of the power ratio measurements are enclosed by these intervals, meaning our measurements are dominated by Poisson noise. However, there are several clusters with power ratios below this level, which implies their diffuse emission is too symmetric compared to pure noise, meaning they are highly regular in shape. 

For use in measuring temperatures and luminosities (see Section \ref{sec:templum}), we also measured and modeled the one-dimensional radial profiles of the diffuse emission. We first measured the counts in circular annuli around the X-ray centroid. For each cluster, we determined at what radius $r_e$ the surface brightness reached the background level, which we used later to define the region for extracting spectra and net counts. We then used the counts in the annuli to fit an azimuthally-averaged surface brightness model consisting of a beta model and a constant background: 
\begin{equation}
SB\left(r\right)=A\left(1+r^2/r^2_c\right)^{1/2-3\beta}+SB_{bkg}
\end{equation}
\label{eq:SB}

As in Equation \ref{eq:betamodel}, $r_c$ is the core radius, and we set $\beta=2/3$. As an additional constraint, we required that the net counts (NC) predicted by the surface brightness model within $r_e$ match the value we measured. This is equivalent to 
\begin{equation}
NC\left(r_e\right)= 2\pi Ar_c^2\left(1 - 1/\sqrt{1 + r_e^2/r_c^2}\right)
\end{equation}
This reduces the independent parameters in our model to two: $r_c$ and $SB_{bkg}$. 

\subsubsection{Temperature and Luminosity}
\label{sec:templum}

To measure the gas temperature of each cluster, we extracted spectra using the CIAO tool {\it specextract}. We extracted the spectra inside circular regions centered on the X-ray centroids, using the $r_e$ values we found in Section \ref{sec:spatprof} as the radii. A background spectrum was extracted nearby and subtracted. 

The spectra were fit to a Raymond-Smith thermal plasma model \citep{RS77}, with the absorption model of \citet{MM83}, which we chose for consistency with previous work. For our models, we fixed $Z=0.3Z_{\odot}$, which is commonly used in the literature \citep{ES91}. Additionally, variations in the metallicity do not have a large impact on the results. We calculated Galactic neutral hydrogen column densities using the COLDEN tool from the {\it Chandra} proposal toolkit, which uses the data set of \citet{DicLock90}. We fit the Raymond-Smith models, using the Sherpa tool from the CIAO toolset, in the energy range 0.5-8 keV, using $\chi^2$\ statistics. Because of low number counts, spectra were grouped to include a minimum of 20 counts per bin. Our results are shown in Table \ref{clusproptab}. Note that we were unable to fit a temperature model to SC1324B with any meaningful precision.  

We measured the net photon counts from each cluster using the same extraction and background regions as the spectra. Since the size of these regions varied from cluster to cluster, we normalized these measurements by extrapolating to $r_{500} \equiv 2\sigma_v /\left[ 500 H\left(z\right)\right]$. We carried out the extrapolation using the surface brightness models we fit in Section \ref{sec:spatprof}. We then used the fitted Raymond-Smith models to convert net counts within $R_{500}$ in the 0.5-2.0 keV range into fluxes in the observer frame in the $0.5/(1+z)$-$2.0/(1+z)$ keV range. Multiplying these values by $4\pi D_L^2$ then converts to the luminosity emitted in the rest frame 0.5-2.0 keV range. The Raymond-Smith models were then used again to extrapolate bolometric luminosities. These values are given in Table \ref{clusproptab}. Since we were unable to measure a temperature for SC1324B, we did not have a Raymond-Smith model to use for extrapolating bolometric luminosities. Since we were unable to measure an accurate X-ray temperature or luminosity, this cluster is excluded from analysis in Section \ref{sec:ana}.

\begin{figure}
    \includegraphics[width=0.45\textwidth]{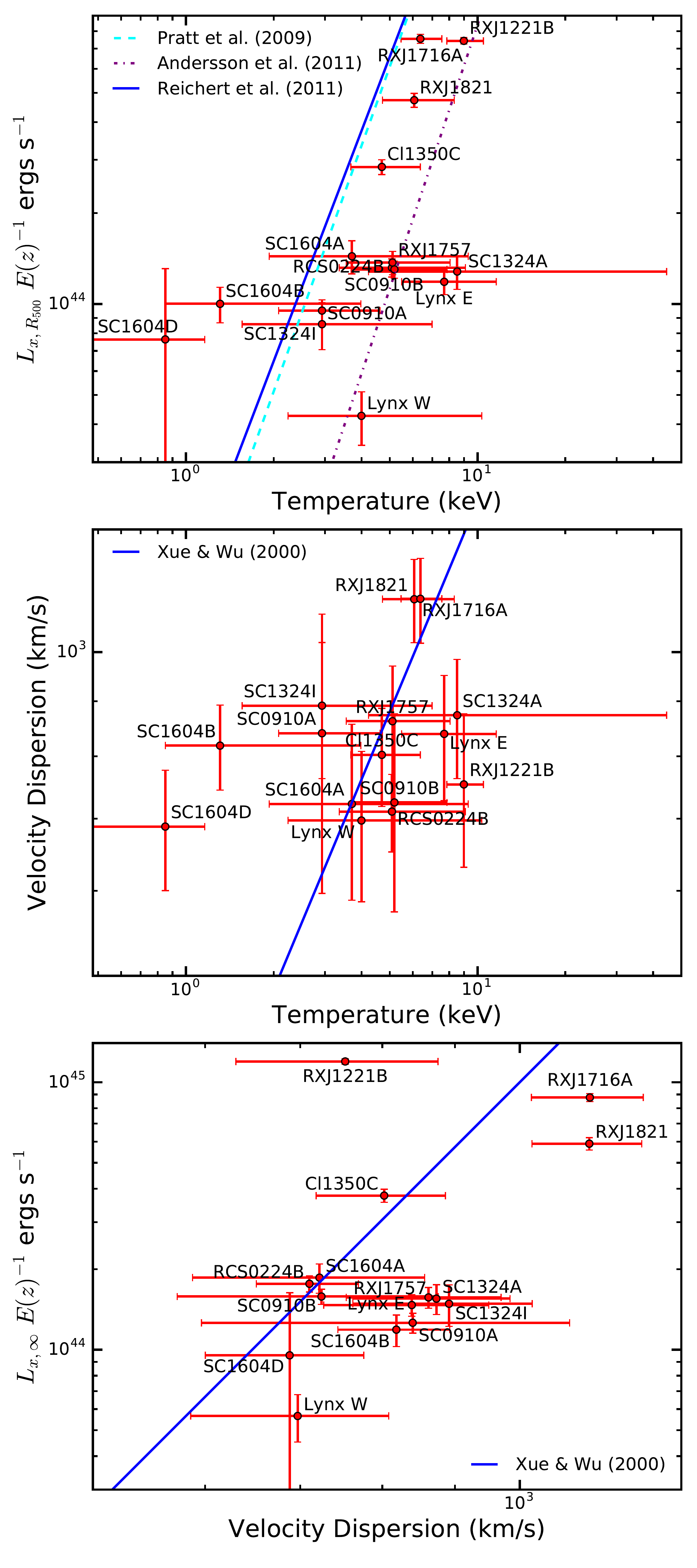}    
    \caption{
Scaling relations between properties of the galaxy cluster. The top plot shows the relation between the bolometric X-ray luminosities and X-ray temperatures of the diffuse gas. Also shown are the scaling relations derived from virialized clusters of \citet{pratt09}, \citet{andersson11}, and \citet{rei11}. In the middle plot, we show the relation between the X-ray gas temperature and the velocity dispersion of the cluster galaxies, along with the scaling relation from \citet{XW00}. In the bottom plot, we show the relation between the bolometric X-ray luminosities and velocity dispersion of the cluster galaxies. Again, we plot the scaling relation from \citet{XW00}. For consistency with \citet{XW00}, the luminosities are extrapolated to infinite radius, rather than $R_{500}$.}
    \label{fig:SR}
\end{figure}

\section{Scaling Relations}
\label{sec:SR}

If a galaxy cluster were subjected only to gravitational interactions, we would expect certain scaling relations between the temperature and luminosity of the ICM gas.  Furthermore, if we are able to effectively remove outliers and select a largely pure and representative member population, these relations should also extend to the line of sight differential velocity dispersions of these galaxies. For example, the only source of heating for the gas, in this case, would be the gravitational collapse of the cluster. With photons emitted via bremsstrahlung emission, we would expect the X-ray luminosity to scale with the gas temperature as $L_x\propto T^2$\ \citep{kaiser86}. We would also expect the coefficient of proportionally to evolve with redshift as $E\left(z\right)$ \citep{krav06}, where 
\begin{equation}
E\left(z\right)=\left[\Omega_m\left(1+z\right)^3+\left(1-\Omega_m-\Omega_{\Lambda}\right)\left(1+z\right)^2+\Omega_{\Lambda}\right]^{1/2}. 
\end{equation}
Similarly, we would expect the velocity dispersion of the cluster galaxies to be related to the ICM temperature as $\sigma_v\propto T^{1/2}$, assuming virialization. It would then follow that $L_x\propto \sigma_v^4$. 

In practice, it has been found that galaxy clusters do not follow these naive scaling relations. Studies at low redshift have found an $L_x$-$T$ relation closer to $L_x\propto T^3$ \citep{mark98, AE99, XW00, vik02}. This steeper relation implies an injection of energy from a non-gravitational source, such as AGN. Similarly, low-redshift studies have found that the $\sigma_v$-$T$ and $L_x$-$\sigma_v$ relations also deviate from the naive expectations, with higher powers of $T$ and $\sigma_v$, respectively \citep{XW00,horner01}, with the same implication of energy from non-gravitational sources. 

In Figure \ref{fig:SR}, we have plotted the ICM temperatures, luminosities, and galaxy velocity dispersions against each other. In the first panel, the bolometric luminosities are used, and we have plotted the local scaling relations found by \citet{pratt09}, \citet{andersson11}, and \citet{rei11}, which follow $L_x\propto T^{2.70}$, $L_x\propto T^{2.90}$, and $L_x\propto T^{2.53}$, respectively. For the other two scaling relations, we plot the empirical relations from \citet{XW00}, which follow $\sigma_v\propto T^{0.65}$ and $L_x\propto \sigma_v^{5.30}$. For the latter, \citet{XW00} use bolometric luminosities extrapolated to infinite radius, using their fitted beta models. Lacking an alternative relation using $L_{500}$, we made the same extrapolation for the $L_x$-$\sigma_v$ relation. 

\begin{table}
\caption{Scaling Relation Offsets}
\label{offsettab}
\begin{tabular}{lcccc}
\toprule
\footnotesize{Cluster}
 & \footnotesize{$L_x$-$T$}
 & \footnotesize{$\sigma_v$-$T$}
 & \footnotesize{$L_x$-$\sigma_v$}
 & \footnotesize{Mean}\\
   {}
 & {offset ($\sigma$)$^{\rm a}$}
 & {offset ($\sigma$)}
 & {offset ($\sigma$)}
 & {offset ($\sigma$)$^{\rm a}$}\\
\midrule
RCS0224B & 1.30(0.05) & 0.83 & 0.48 & 0.87(0.45) \\
SC0849C & 2.26(1.18) & 1.27 & 1.52 & 1.68(1.32) \\
SC0849D & 1.21(0.23) & 0.26 & 1.29 & 0.92(0.84) \\
SC0910A & 0.49(1.05) & 0.66 & 0.81 & 0.65(0.84) \\
SC0910B & 1.64(0.03) & 0.78 & 0.29 & 0.90(0.37) \\
RXJ1221B & 3.18(0.41) & 3.23 & 1.56 & 2.66(1.74) \\
SC1324A & 1.33(0.76) & 0.81 & 1.74 & 1.29(1.10) \\
SC1324I & 0.37(0.40) & 0.52 & 2.01 & 0.97(0.98) \\
Cl1350C & 0.91(1.31) & 1.15 & 0.16 & 0.45(0.59) \\
SC1604A & 0.44(0.31) & 0.01 & 0.21 & 0.22(0.18) \\
SC1604B & 0.47(1.31) & 1.15 & 2.25 & 1.29(1.57) \\
SC1604D & 1.30(1.39) & 3.86 & 0.63 & 1.93(1.96) \\
RXJ1716A & 1.06(2.76) & 0.57 & 1.69 & 1.11(1.68) \\
RXJ1757 & 1.43(0.08) & 0.14 & 1.61 & 1.06(0.61) \\
RXJ1821 & 1.12(0.95) & 0.48 & 2.51 & 1.37(1.31) \\
\bottomrule
\multicolumn{5}{p{8.5cm}}{$^{\rm a}$ \footnotesize{Offsets from the $L_x$-$T$ use the \citet{rei11} relation. Values in parentheses use the \citet{andersson11} relation instead. See Section \ref{sec:SR} for details.}}
\end{tabular}
\end{table}

In Table \ref{offsettab}, we have compiled the offsets from the scaling relations for each cluster, normalized by the uncertainties on the measurements\footnote{Scaling relation offsets are calculated as the shortest distance to the scaling relation curve.}. Since there is substantial variation in the $L_x$-$T$ relations from the literature\footnote{Note that the variation between scaling relations is within the uncertainties on their parameters.}, as seen in Figure \ref{fig:SR}, we recorded offsets relative to both the \citet{rei11} and \citet{andersson11}, with the latter in parentheses. The \citet{rei11} relation is derived from a large sample compiled from the literature, with a wide range of redshifts ($z<1.1$). The \citet{andersson11} sample is derived from a smaller sample, but with a similar redshift range to the ORELSE survey ($0.4<z<1.1$). Because of a relative lack of research into the $\sigma_v$-$T$ and $L_x$-$\sigma_v$ relations, the \citet{XW00} relations are the only ones available to us, and we therefore cannot compare different derived relations in these cases. In addition, we lack the dynamic range necessary to fit meaningful relations ourselves given our sample size. 

\section{Analysis of Virialization Tests and Correlations to Scaling Relation Offsets}
\label{sec:ana}

We have compiled a set of tests that have resulted in a set of measurements designed to probe the virialization and presence of substructure in galaxy clusters. In addition, we computed the velocity dispersions, X-ray temperatures, and luminosities of all clusters in our sample. Comparing these values to relations between virialized clusters, we would expect offsets from these relations to be correlated with the results of our virialization/substructure tests. By examining the strength of correlations, we can determine which tests have the most power in predicting the virialization of galaxy clusters. 

We stress that in the analysis that follows we rely completely on the assumption that clusters with small offsets from the adopted empirical scaling relations are near virialized, while those exhibiting large offsets relative to these relations are not. We further stress that the purpose of this exercise is not to definitively determine which ORELSE clusters are in a near virialized (or unrelaxed) state or at what value of a certain metric a cluster can be considered near virialized, but rather to understand which of the measurements\footnote{Here we refer only to those measurements in Section \ref{sec:CP} that do not explicitly go into calculating scaling relation offsets, i.e., not $L_{X}$, $T_{X}$, or $\sigma_{v}$.} described in Section \ref{sec:CP} correlate well with scaling relation offsets measured in Section \ref{sec:SR} in order to inform observational strategies for general cluster surveys. The first step in the process of this testing is to gather all of our tests of virialization and substructure and to normalize them so that they may properly compared in a common framework. For brevity, we will refer to these normalized measurements as ``metrics".

\subsection{Normalizing Measurements}
\label{sec:norm}

First, in Section \ref{sec:veldisp}, we examined the velocities of different populations of galaxies within clusters. We derive two metrics: the difference between the velocity centers of the red/quiescent and blue/star-forming populations, and the differences between their velocity dispersions. For the difference in velocity dispersions, the metric was normalized by dividing the difference by its uncertainty, effectively converting our units to the uncertainty in the velocity dispersion, $\sigma$, under the assumption of a Gaussian distribution. For the difference in velocity centers, we calculated the probability that the difference arose by chance, using Monte Carlo simulations where the colours or star-forming statuses of the galaxies were randomly shuffled among the sample in each trial (see Section \ref{sec:RB}). This was normalized by assuming a Gaussian distribution as well. $P\left(\Delta_v\right)$ is our measure of confidence that $\Delta_v$ arose by chance, which can be thought of in units of $\sigma$. For example, a 32\% chance of measuring a velocity center difference at least as high as observed would correspond to $\sigma=1$, and to have a confidence of $3\sigma$, $P\left(\Delta_v\right)$ would have to be $\le0.3$\%. 

Next, we performed the DS tests of substructure in Section \ref{sec:DStest}. The test outputs the parameter $\Delta$, whose distribution depends on the specific coordinates and redshifts of the galaxies involved. We used Monte Carlo simulations to obtain a distribution of $\Delta$ values for each cluster. We then calculated $P\left(\Delta\right)$, the chance of measuring a $\Delta$ value at least as high as observed by observing how many Monte Carlo trials had a $\Delta$ value at least that high. We can normalize this metric in the same way we normalized the velocity center differences above: we converted $P\left(\Delta\right)$ to units of $\sigma$ by assuming a Gaussian distribution. $P\left(\Delta\right)$ is our measure of confidence, so $1\sigma$ would correspond to $P\left(\Delta\right)=32$\%, $2\sigma$ would correspond to $P\left(\Delta\right)=4.6$\%, etc.

We then examined the asymmetry of the diffuse X-ray emission in Section \ref{sec:DE}. To quantify the asymmetry, we measured the power ratios $P_3/P_0$ and $P_4/P_0$. These measurements were especially noisy, with most of the variation explained by noise. There were, however, some clusters with power ratios lower than would be expected by Poisson noise in the X-ray images. This evidence suggests these are the clusters with the most symmetric ICM emission, which provides some information. We encode this information into a normalized metric by assigning clusters with these especially low power ratios measurements a metric value of -1, while assigning a value of zero to all others. 

We calculated three different galaxy cluster centroids: the centers of X-ray emission, the MMCGs/BCGs, and weighted centers using luminosities/masses (see Sections \ref{sec:WMC}, \ref{sec:MMCG}, and \ref{sec:DE}). The distances between these measurements are three different measurements of virialization, although only two are independent. They were normalized by dividing the distances by their uncertainties, using errors on the positions. 

Two different methods were combined to estimate the uncertainty for the weighted mean centers. In the first method we measured the offset between the LWMCs/MWMCs of each cluster calculated by spec-z members only and those centers calculated from a combination of spec-$z$ members and photo-$z$ members which did not have a secure spectroscopic redshift outside of the cluster bounds (see Section \ref{sec:MMCG2} for details on photo-$z$ membership criteria). This method was used to account for uncertainties related to the incompleteness of our spectroscopic campaign. In the second method we carried out Monte Carlo simulations to estimate the error associated with choosing a specific radius around the X-ray centroid within which to measure the weighted mean centers. In each trial, the radial cut was randomly varied between 0.5 and 1.25 $h_{70}^{-1}$\ Mpc and 5\% of the spectroscopic sample was cut. As an ensemble, variations in the weighted mean centers were found to be $<0.1h_{70}^{-1}$\ Mpc for 84\% of trials, for both luminosity and mass weighting. The median value offset value from these simulations ($\sim$$0.05h_{70}^{-1}$\ Mpc) was added in quadrature with the uncertainties from the first method to estimate the total uncertainty on the weighted mean centers.

\begin{table*}
\caption{Virialization Metric Correlations}
\label{corrtab}
\begin{tabular}{lcccc}
\toprule
\footnotesize{Metric}
& \multicolumn{4}{c}{\footnotesize{$R^2$ without Metric$^{\rm a}$}} \\
{} & \multicolumn{2}{c}{\footnotesize{Reichert et al. (2011)$^{\rm b}$}} & \multicolumn{2}{c}{\footnotesize{Andersson et al. (2011)$^{\rm b}$}} \\
   {}
 & \footnotesize{(SED fits)}
 & \footnotesize{(no SED fits)}
 & \footnotesize{(SED fits)}
 & \footnotesize{(no SED fits)}\\
\midrule
Galaxy populations velocity center offset$^{\rm c}$ & {\bf 0.204 (0.123)} & {\bf 0.496 (0.132)} & {\bf 0.412 (0.226)} & 0.442 (0.006)\\
Galaxy populations velocity dispersion diff.$^{\rm c}$ & 0.305 (0.021) & 0.618 (0.009) & 0.465 (0.172) & 0.433 (0.015)\\
DS test & 0.306 (0.020) & 0.595 (0.033) & 0.540 (0.098) & 0.441 (0.007)\\
$P_3/P_0$ & 0.291 (0.035) & {\bf 0.501 (0.127)} & {\bf 0.446 (0.191)} & {\bf 0.330 (0.117)}\\
$P_4/P_0$ & 0.325 (0.001) & 0.618 (0.010) & 0.628 (0.010) & 0.439 (0.009)\\
MMCG/BCG to X-ray distance$^{\rm d}$ & 0.295 (0.031) & {\bf 0.542 (0.086)} & 0.484 (0.153) & 0.448 (0.000)\\
MMCG/BCG to MWMC/LWMC distance$^{\rm d,e}$ & {\bf 0.258 (0.068)} & 0.619 (0.009) &  {\bf 0.461 (0.177)} & {\bf 0.416 (0.032)}\\
MWMC/LWMC to X-ray distance$^{\rm e}$ &  {\bf 0.231 (0.096)} & {\bf 0.082 (0.546)} & {\bf 0.252 (0.385)} & {\bf 0.270 (0.178)}\\
MMCG/BCG Velocity & 0.307 (0.020) & 0.553 (0.075) & 0.509 (0.129) & {\bf 0.408 (0.039)}\\
Quiescent fraction & {\bf 0.252 (0.075)} & 0.000 (0.628) & 0.467 (0.171) & 0.000 (0.448)\\
\hline
Total $R^2$ & 0.327\phantom{ (0.021)} & 0.628\phantom{ (0.021)} & 0.638\phantom{ (0.021)} & 0.448\phantom{ (0.021)} \\
\bottomrule
\multicolumn{5}{p{16cm}}{$^{\rm a}$ \footnotesize{Results show the goodness-of-fit metric, $R^2$, when performing ordinary least squares regression between the combined virialization metrics listed and the mean offset from the scaling relations. The Total $R^2$ row uses all metrics, while the other rows use all but the listed metric. Values in parentheses are the $R^2$ calculated with the exclusion of the relevant metric subtracted from the Total $R^2$. The four highest reductions in $R^2$ in each column are shown in bold.}}\\
\multicolumn{5}{p{16cm}}{$^{\rm b}$ \footnotesize{Results use either the \citet{rei11} or \citet{andersson11} $L_x$-$T$ scaling relation. See Section \ref{sec:SR} for details.}}\\
\multicolumn{5}{p{16cm}}{$^{\rm c}$ \footnotesize{Difference between between quiescent and star-forming populations for test using SED fitting metrics, and difference between blue and red populations for test without SED fitting.}}\\
\multicolumn{5}{l}{$^{\rm d}$ \footnotesize{Distance measure uses MMCG for test using SED fitting metrics, and BCG for test without SED fitting.}}\\
\multicolumn{5}{l}{$^{\rm e}$ \footnotesize{Distance measure uses MWMC for test using SED fitting metrics, and LWMC for test without SED fitting.}}\\
\multicolumn{5}{l}{$^{\rm f}$ \footnotesize{Since SED fitting was required for estimating the quiescent fraction, it was not used for these linear fits.}}
\end{tabular}
\end{table*}

\begin{figure*}
    \includegraphics[width=0.99\textwidth]{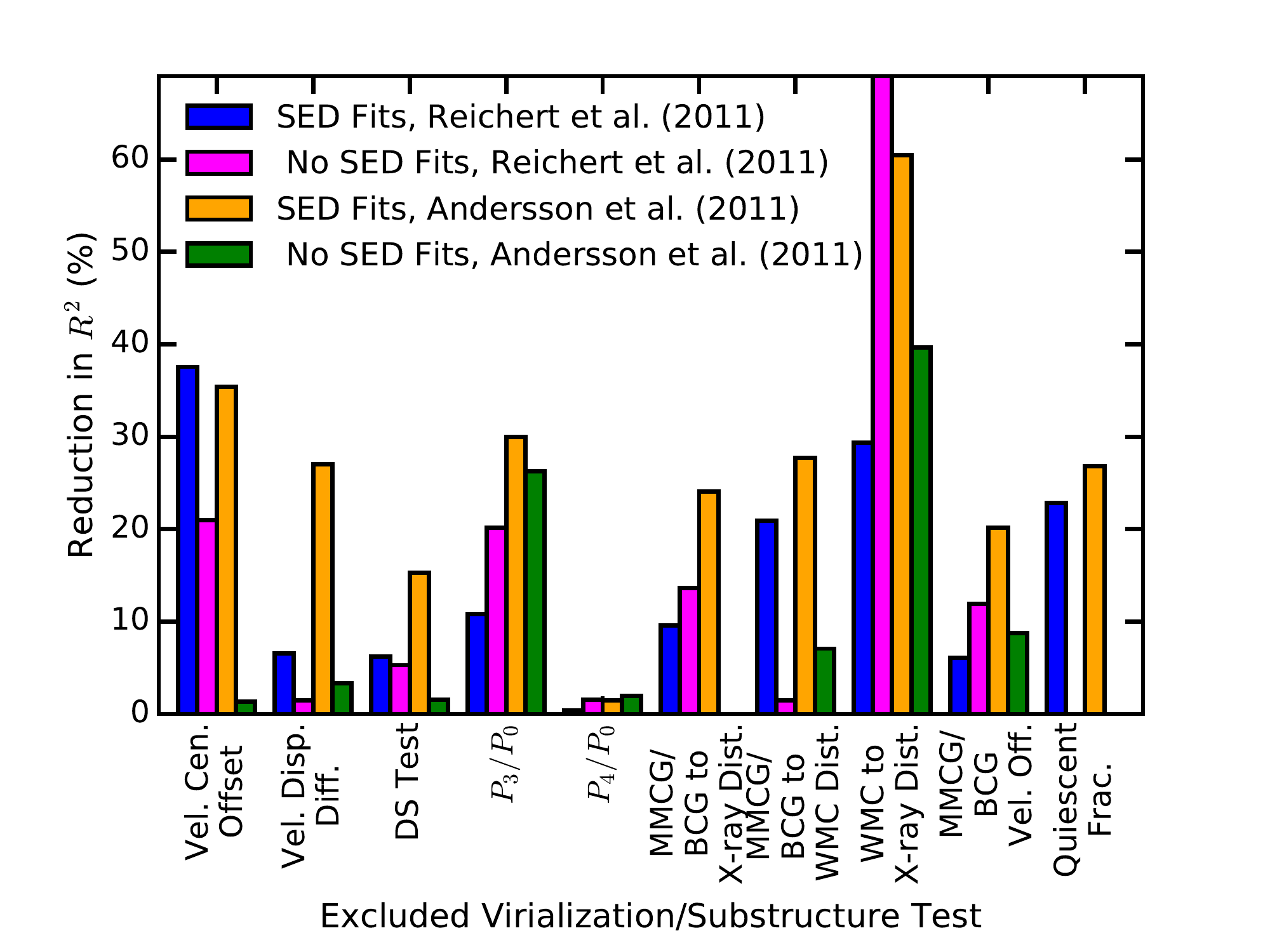}
    \caption{
Results from performing ordinary least squares regression between the combined virialization metrics and the mean offset from the scaling relations. Plotted are the percentage reductions in the goodness-of-fit metric, $R^2$, when all metrics except the one listed are used, compared to when every metric is used.}
    \label{fig:metrics}
\end{figure*}

Uncertainties on the X-ray centroids were estimated using Monte Carlo simulations performed as follows. For each cluster, we took a cutout of the unsmoothed image around the cluster and measured the surface brightness at each point, smoothing with a tophat kernel because the average counts per pixel was less than 1. For each Monte Carlo trial, we randomized the photon counts for each pixel using a Poisson distribution with the surface brightness as the estimate of the expected value, then re-smoothed the simulated image and located the X-ray center. The distribution of simulated X-ray centers gives us an estimate of the uncertainty on the X-ray centroid. 

Compared to the uncertainties on the mean centers and X-ray centroid, the positional uncertainty of the galaxy identified as the MMCG/BCG is negligible. While there is a systematic uncertainty associated with the potential misidentification of the MMCG/BCG, we definitively identify each in all clusters, so such a concern does not apply to this work. The three distances between centroids are therefore normalized by the uncertainties on the weighted mean center and X-ray centroid, adding them in quadrature for the distance between these two centroids, and using only the single relevant uncertainty when normalizing the offsets from the MMCG/BCG. The distance between the MMCG/BCG and the cluster center in velocity space also provides information on the virialization of a cluster. The uncertainty on the velocity offset should, in principle, be measured from a combination of the uncertainty on the redshift of the MMCG/BCG, the uncertainty in the systemic velocity, and the uncertainty relating to whether a given velocity is meaningful relative to the random motions in the cluster (i.e., relative to $\sigma_{v}$). In practice, however, the uncertainty in redshift of the MMCG/BCG is negligible ($\sim10-20$ km s$^{-1}$) and the uncertainty in the systemic redshift of a given cluster based on bootstrap estimates\footnote{A better way to approach the estimate of this uncertainty would be to use an identical method to what was used to estimate the LWMC/MWMC uncertainties. However, the lack of velocity precision of the photo-$z$ measurements prevents such a method from being used.} is also negligible ($\sim100-200$ km s$^{-1}$) to the precision necessary for the calculation here. Thus, to normalize this metric we simply divide the velocity offsets by the velocity dispersion of each cluster.

Lastly, we have the quiescent fractions, described in Section \ref{sec:SF}. We may expect more virialized clusters to also have higher quiescent fractions, since their members have spent more time near the cluster core, where process that quench star formation are stronger. However, we do not need to normalize these measurements, since they are already directly comparable between clusters, and we expect them to correlate directly with the degree of virialization.

\begin{figure*}
    \includegraphics[width=0.68\textwidth]{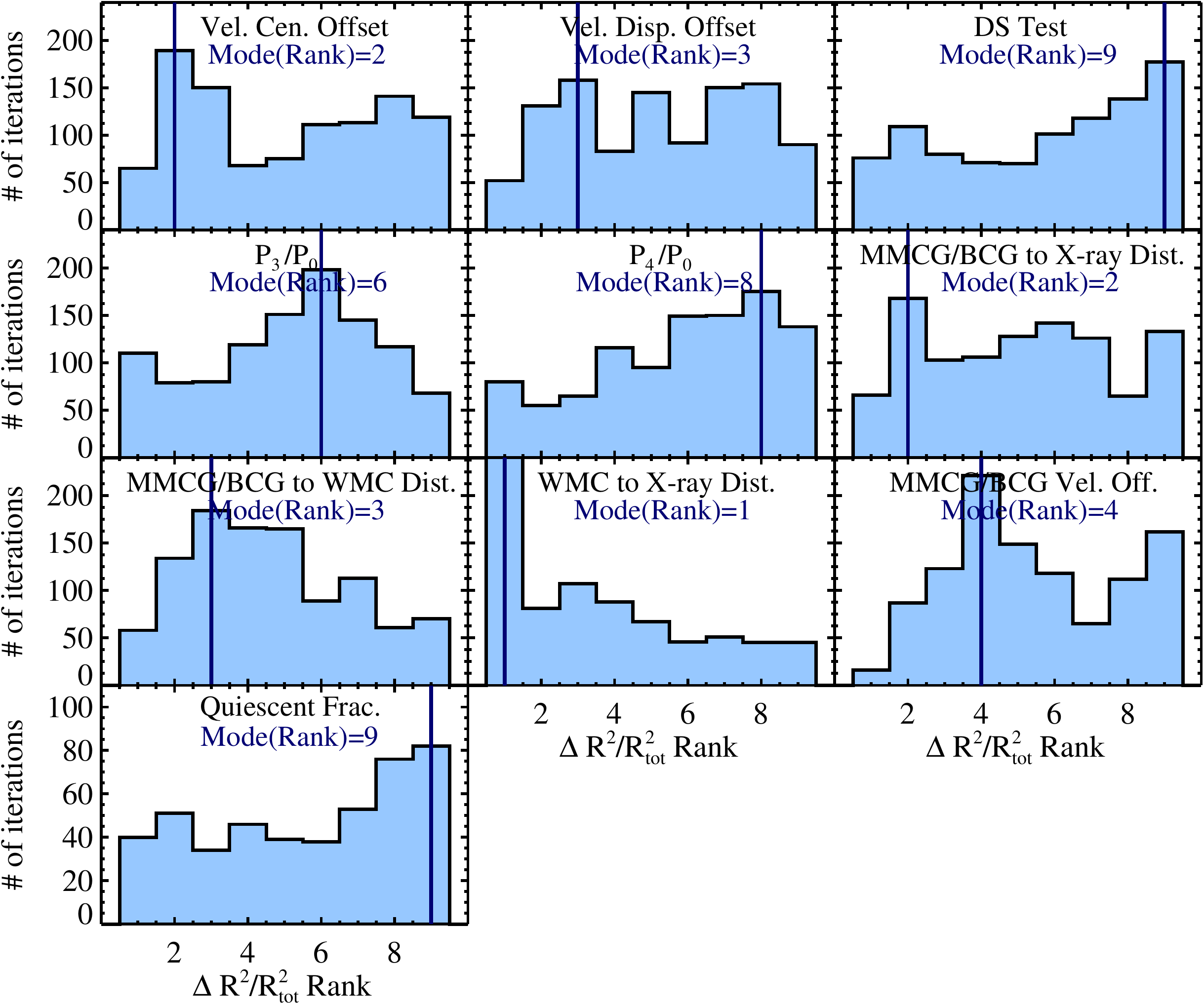}
    \includegraphics[width=0.68\textwidth]{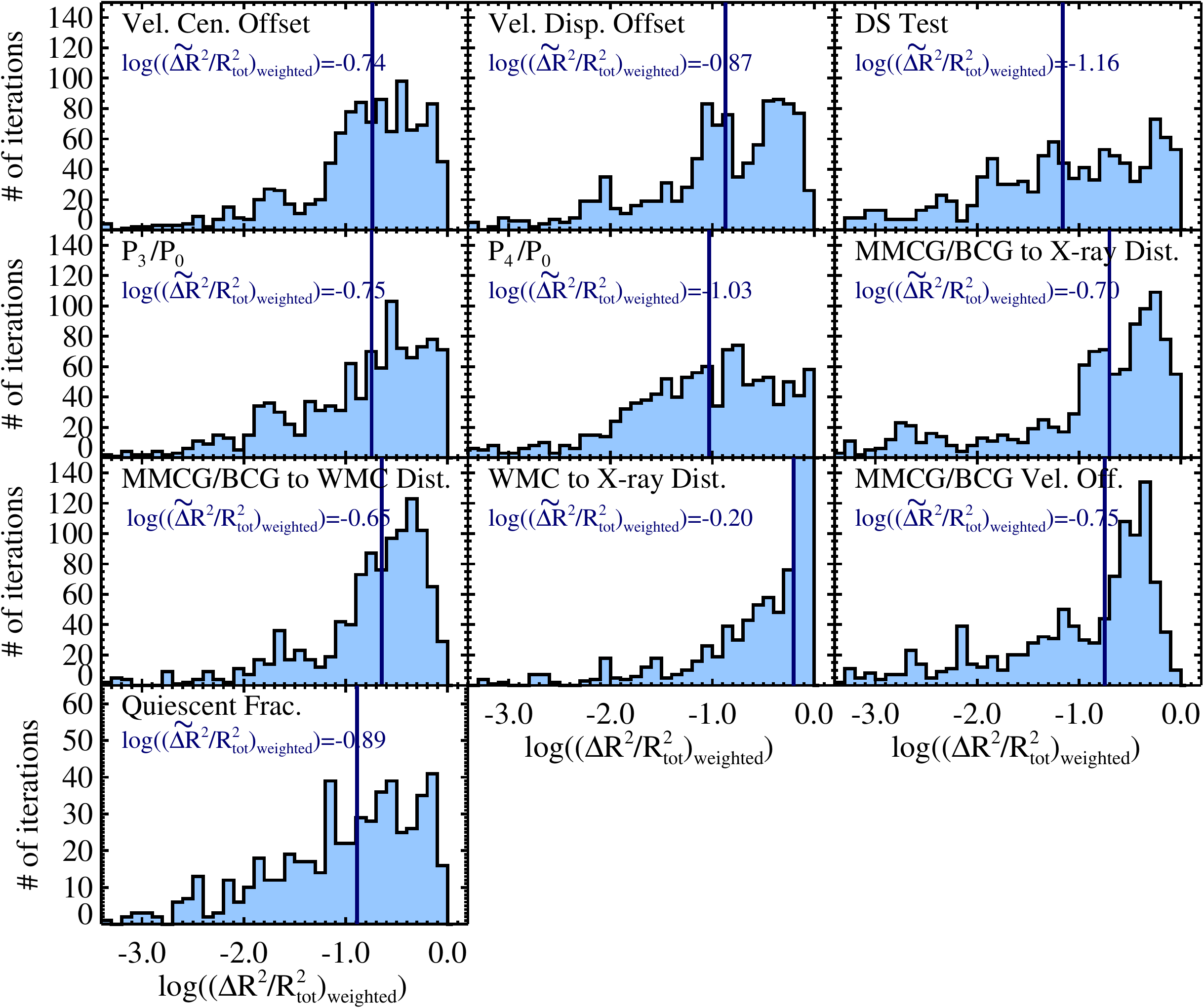}
    \caption{
\emph{Top:} The distribution of the ranked importance of each metric in reducing $R^2$ when removed from the joint fit relative to the scaling relation offsets for 
different sub-samples of the ORELSE clusters, metrics which use and do not use SED fitting, different normalization approaches, and offsets with respect to different 
scaling relations (see Section \ref{sec:predict} for details). Lower numbers indicate a metric is relatively more important in predicting cluster virialization. The mode 
of each distribution is shown by the solid line and the value is given in each sub-panel. For consistency, the 10th most important metric for those with iterations which 
employ SED fitting was not considered in these distributions$^{11}$. \emph{Bottom:} The distribution of the weighted percentage reductions in $R^2$ (see Section \ref{sec:predict})
for each metric for the same cases as were considered in the left panels. The median of each distribution is shown by the solid line and the value is given in each sub-panel. Higher numbers indicate a metric is relatively more important in predicting cluster virialization.} 
    \label{fig:metricdist}
\end{figure*}

\subsection{Metric Correlations with Scaling Relation Offsets}
\label{sec:corr}

We may expect each of the normalized metrics described in the previous section to correlate with the offsets from the scaling relations. We can test this by performing linear regression between the metrics and scaling relation offsets. 

For each metric, we used the \textsc{LinearRegression} class from {\it scikit-learn} to perform ordinary least squares regression with the offsets from each of the scaling relations, performed for both the \citet{andersson11} and \citet{rei11} relations in the case of the $L_x$-$T$ relation, as well as the mean offset from all three scaling relations again using those offsets from the two $L_x$-$T$ relations separately. To evaluate the goodness of fit, we calculated $R^2$ for each case. In every case, $R^2$ was less than 0.37, indicating weak to moderate correlations between individual metrics and both individual and mean offsets from the various scaling relations. 

The correlation between the metrics and the scaling relation offsets can be improved by fitting them jointly. In short, we want to find the coefficients such that $\Sigma_i \alpha_i m_i$ is a good predictor of the offset from the scaling relations, where the $m_i$ are the different metrics and the $\alpha_i$ are constant coefficients. We did this using two sets of metrics: one using metrics derived using SED fitting, and one using metrics without using SED fitting, the latter omitting photometric redshifts, rest-frame colours, stellar masses, and quiescent fractions. The rationale here is that the parameters derived from SEDs may be more accurate indicators of certain properties (e.g., the masses derived from SED fits can more accurately locate the MMCG, as opposed to using the luminosities as a proxy for mass with the BCG), but these may not be available for all surveys. Metrics such as LWMCs are much less expensive to calculate, in terms of observation and computation, than MWMCs. So, a set of metrics with metrics derived from SED fitting may provide insights into how parameters correlate with virialization when one has the luxury of accurately and precisely estimating SEDs, while the second set of metrics provides a comparison for when this is not feasible.

In addition, we measured offsets from the $L_x$-$T$ scaling relation using the fitted relations of both \citet{rei11} and \citet{andersson11}, and we carry out the analysis using both sets of offsets for comparison. With the parallel analyses using and not using the SED fitting, this makes four separate analyses, as shown in Table \ref{corrtab}.

For each set of metrics, we performed a linear fit and estimated the goodness of fit using $R^2$, again using {\it scikit-learn} as described above. In this case, the goodness of fit metric is calculated as \begin{equation}
R^2=1-\frac{\Sigma_i \left(y_{i,obs}-y_{i,pred}\right)^2}{\Sigma_i \left(y_{i,obs}-\left<y_{i,obs}\right>\right)^2}
\end{equation}
The $R^2$ values are given in the row marked ``Total'' in Table \ref{corrtab}. The values of $R^2$ range from 0.33-0.64, a marked improvement over individual fits.

While the joint fits show that a suite of virialization tests can be effective in predicting offsets from scaling relations, we would like to investigate the individual power of the tests. We actually can do so using the joint linear fits. If an individual metric provides meaningful information for predicting scaling relation offsets, removing it from the set of metrics and reevaluating the linear fits without it should reduce the Total $R^2$ (hereafter simply $R^2$). The larger the decrease in $R^2$, the more predictive power that metric provided. The relative sizes of the decreases in $R^2$ should also tell us the relative importance of the metrics. We carry out this test for each metric described in Section \ref{sec:norm}. The $R^2$ calculated from the revised fits with one metric removed are given in Table \ref{corrtab}, with the reduction in $R^2$ in parentheses. The percentage reductions compared to the $R^2$ are also plotted in Figure \ref{fig:metrics}. The reductions in $R^2$ range from almost negligible to more than 0.8 (i.e., over 80\%) indicating a wide range of influences from the different metrics. Broadly, the spatial offsets between the various projected centers (MMCG/BCG, WMC, X-ray) as well as the differences in the mean and dispersion of the velocities of red and blue member galaxies and the X-ray power ratio P$_3$/P$_0$ appear to have the most predictive power with regards to the scaling relation offsets of the ORELSE clusters irrespective of the framework of the analysis (i.e., SED fits vs. no SED fits, scaling relation used). Nearly all six of these metrics appear among the top four most important metrics in terms of $R^2$ reduction in Table \ref{corrtab} in at least one framework and usually more than one. Conversely, the X-ray power ratio P$_4$/P$_0$, the DS tests, and the quiescent fraction appear to have limited predictive power. These three metrics rarely appear among the top four largest reductions in $R^2$. While it is possible that surface brightness issues contribute to the limited combined predictive power of the two power ratios, it is unlikely as both $P_3/P_0$ and $P_4/P_0$ have been shown to be sensitive to sub-structure for clusters of similar masses and redshifts to those of our own sample using similar observational setups \citep{jel05}.

\subsection{General Predictive Power of Metrics}
\label{sec:predict}

In the previous section we established that some metrics appear to have more power than others in predicting scaling relation offsets and, by consequence, the virialization state of ORELSE clusters. Though the number of clusters studied here is relatively modest, the sample is selected through a variety of different techniques (see \citealt{lubin09}) and, further, spans a large range in dynamical mass ($\log(\mathcal{M}_{vir}/\mathcal{M_{\odot}}) = 14.4-15.1$ using the methodology of \citealt{lem12}). As such, the results derived here should be useful in informing observing strategies for future surveys if, indeed, these results can be extended to the general population of intermediate redshift galaxy clusters. To this end, we attempted to test the robustness of these results to changes in both the sample and the methodology by following the approach in Section \ref{sec:corr}, calculating the reduction in $R^2$ when removing each metric (hereafter $\Delta R^2/R^2$), for a large suite of conditions. These conditions included different samples, achieved through jackknifing the ORELSE cluster sample, using offsets with respect to all three different scaling relations individually, and changes in the approach used to estimate uncertainties, done through replacing the fiducial velocity dispersion and WMC uncertainties with estimates using a bootstrap approach. In this way we created $\sim$1000 unique values of $\Delta R^2/R^2$ for each metric\footnote{Since the quiescent fraction was only calculated for iterations which included SED fitting, there are only $\sim$500 iterations for this metric.} that span an enormous range of combinations of measured quantities (SED fitting vs. no SED fitting), different scaling relations against which to measure offsets, different cluster samples, and different approaches in the way some of the metrics are normalized. Though in the latter case the approach is changed for only two of the metrics, $\Delta R^2/R^2$ for a given iteration depends on all metrics in that iteration resulting in a unique value being generated for each metric for each iteration. Because of the immense range of conditions these different iterations cover, if certain metrics appear to consistently return higher $\Delta R^2/R^2$ when removed from the fits across all iterations it is likely these metrics would be useful in predicting the virialization state of clusters for a general cluster survey. 

The top panel of Figure \ref{fig:metricdist} shows the distribution of the ranked importance of each metric in terms of $\Delta R^2/R^2$ for 
all iterations along with the mode of each distribution. Our convention here is to assign a rank of one to the most important metrics for a given
iteration and a rank of nine to the least important\footnote{Note that, since those iterations that did not include metrics 
which employed SED fitting had a maximum rank of nine, the quiescent fraction metric being removed, the tenth ranked metric for each iteration which employed 
SED fitting was not considered on this plot.}. While the distribution of the ranked importance of 
each metric is broad, spanning from most important to least important in all cases, it is clear from a visual inspection that the distributions are considerably 
different in shape. In some cases
the distribution is skewed towards higher ranks while others appear be generally less highly ranked or have a near uniform distribution. The two metrics that 
appeared to be consistently the most important were the projected offset between the WMC and X-ray centers and the projected offset between the 
location of the MMCG/BCG and the WMC. These two metrics appeared in the top three most important metrics 68.8\% and 36.5\% of the time, respectively. In contrast, 		
the DS test, the quiescent fraction, and the two power ratios appeared to consistently have minimal predictive power, appearing in the top three most important metrics a combined 
24.6\% of the time. These results		
are bolstered by the results of KS tests run on the various distributions. According to these tests, the distribution of the ranks of the two most 
important metrics are the only two metrics which are statistically distinguishable at the $>>$3$\sigma$ level from the distributions of all 
other metrics. 

However, to simply rank order the importance of each metric in predicting the virialization state of a given cluster sample is perhaps an overly simplistic 
approach as it does not consider the magnitude of the reduction in the $R^2$ for the removal of a given metric for a given iteration. As can be seen in Table
\ref{corrtab}, lower ranked metrics, in some cases, still exhibit markedly high $\Delta R^2/R^2$ values, while, in other cases, the
reductions of similarly ranked metrics are essentially negligible. Additionally, there are some cases where a single metric
or a few metrics have dominate $\Delta R^2/R^2$ values 
and other cases in which the distribution is more egalitarian (see the numbers for the no SED fits vs. the \citealt{rei11} relation and the SED fit metrics vs. 
the \citealt{andersson11} relation in Table \ref{corrtab} for an example of each, respectively). 
In order to avoid conflating these cases, we defined a weighted $R^2$ reduction, $(\Delta R^2/R^2)_{weighted}$, for the $i$th metric and the 
$j$th iteration as:

\begin{multline}
(\Delta R^2/R^2)_{weighted,i,j}= \\ 
\frac{(\frac{\Delta R^2}{R^2})_{i}}{\frac{\sum_{i=1}^{n}(\frac{\Delta R^2}{R^2})_{i}-(\frac{\Delta R^2}{R^2})_{1}-(\frac{\Delta R^2}{R^2})_{n}}{n}+(\frac{\Delta R^2}{R^2})_{1}}
\end{multline}
 
\noindent where $n$ is the number of metrics used for the $j$th iteration, i.e., 9 or 10 for no SED fit and SED fit cases, respectively, and 
$(\Delta R^2/R^2)_{1}$ and $(\Delta R^2/R^2)_{n}$ are the percentage reduction of the first and last most important metric for this iteration. 
The quantity $(\Delta R^2/R^2)_{weighted}$ is designed such that all values are between zero and unity, zero for small reductions relative to the most important 
metric and unity for a single, dominate metric. Egalitarian cases return a value of 0.5 for all metrics. 

The bottom panel of Figure \ref{fig:metricdist}
shows the distribution of $(\Delta R^2/R^2)_{weighted}$ for all iterations for each metric along with the median value. We plot $\log(\Delta R^2/R^2)_{weighted}$
rather than $(\Delta R^2/R^2)_{weighted}$ to visually highlight the differences between the various distributions. As was the case for the ranked order
distributions, the projected offset between the WMC and X-ray centers as well as the projected offset between the WMC and the MMCG/BCG 
appear to consistently have the highest values of $(\Delta R^2/R^2)_{weighted}$. These two metrics have the highest median $(\Delta R^2/R^2)_{weighted}$ values
and appear in the top half of all $(\Delta R^2/R^2)_{weighted}$ values a combined 68.0\% of the time. Again the quiescent fraction, power ratios, with the 
possible exception of P$_3$/P$_0$, and DS test values appear to be have generally lower median values and only appear in the top half of the full 
$(\Delta R^2/R^2)_{weighted}$ distribution a combined 40.2\% of the time. In other words, the combination of the WMC to MMCG/BCG-WMC Distance and the WMC
to X-ray Distance metrics was more predictive of the virialization state of a given cluster sample approximately twice as frequently as any combination of two of metrics
with generally lower predictive power (i.e., the quiescent fraction, power ratios, and DS tests), and their median $(\Delta R^2/R^2)_{weighted}$ is nearly four times 
times higher. 		
Taken alone, the WMC to X-ray Distance is unequivocally the most important of all metrics, with a median $(\Delta R^2/R^2)_{weighted}$ that is more than twice 
that of any other metric and is nearly an order of magnitude more predictive on average than the lowest rank metric (DS test). 

Also mirroring the results on ranked importance, KS tests on the $(\Delta R^2/R^2)_{weighted}$ distributions of the two most important metrics find them statistically 
distinguishable from those of every other metric at the $>>$3$\sigma$. In order of importance the remaining metrics are: the projected offset between the 
BCG/MMCG and the X-ray center (falling 53.3\% of the time in the top half of the $(\Delta R^2/R^2)_{weighted}$ distribution), the offset in the velocity center of
red and blue member galaxies (51.3\%), the offset between the MMCG/BCG velocity and the systemic velocity of its parent cluster (50.1\%), and the velocity dispersion 
offset between 
red and blue member galaxies (43.8\%). While the gradations in importance might seem small between the various metrics, the fact that we have tested a huge
variety of circumstances and seen some metrics clearly maintain predictive power and some consistently lack predictive power is telling. In the next
section we summarize these results and suggest observing strategies for different types of surveys based on these results.

\section{Discussion}
\label{sec:disc}

As part of the ORELSE survey, we searched for diffuse X-ray emission in 12 LSSs, finding emission from a total of 16 galaxy clusters. In Section \ref{sec:CP}, we studied the properties of these galaxy clusters, and performed a number of tests of virialization and substructure on them, using a separate set of tests that both did and did not use SED fitting, to allow generalization to a wide range of studies and observational datasets. These galaxy cluster properties and tests are given in Tables \ref{clusproptab}, \ref{centab}, \ref{sigtab}, and \ref{testtab}. 

We would expect virialized clusters to follow relations between properties such as the temperature and luminosity of their diffuse gas, as well as the velocity dispersions of their galaxies, such as those plotted in Figure \ref{fig:SR}. We would expect a galaxy cluster that is still in the process of forming, or that has been recently disrupted by an interaction such as a merger, to be offset from these relations. We find varying degrees of offset from the scaling relations, as shown in Table \ref{offsettab}, which we take to indicate that our sample includes both near virialized and non-virialized clusters. Note that, because a wide range of $L_x$-$T$ relations exist in the literature, we evaluate the offsets from two different $L_x$-$T$ relations: \citet{rei11} and \citet{andersson11}. For the purposes of the analysis presented in this paper, we adopt offsets from the various scaling relations as our proxy for the degree of virialization of a given cluster, with smaller offsets implying a higher level of virialization.

As discussed in Section \ref{sec:corr}, these offsets from the scaling relation should then correlate, to varying degrees, with the virialization and substructure tests we performed. However, correlation between individual metrics and both the offsets from individual scaling relations and the mean offset from all scaling relations were relatively weak, meaning any individual test is, by itself, insufficient as a predictor of virialization. When all metrics given in Table \ref{corrtab} were combined for a linear fit to the mean scaling relation offset, the correlation was relatively strong ($R^2$ ranged from $\sim$ 0.33$-$0.64). Performing every test may be prohibitively expensive, however, involving potentially years of both observation and effort over a wide range of wavelengths, as was the case for the ORELSE survey. 

It is useful, then, to determine which metrics have the most predictive power when it comes to testing cluster virialization and substructure, using scaling relation offsets as a proxy. Rather than using linear fits of individual metrics to the scaling relations, we used the joint fits, as described in Section \ref{sec:corr}. By removing one metric from our set and re-performing the linear fit to the scaling relation offsets, we can evaluate how informative that metric was through the corresponding drop in $R^2$ after excising it. The results of this exercise are shown in Table \ref{corrtab}, with different results for the set of metrics that do and do not use SED fitting, and for offsets from the \citet{rei11} versus \citet{andersson11} $L_x$-$T$ relations. This line of analysis was expanded on in Section \ref{sec:predict} to incorporate a variety of different sub-samples, different scaling relation offsets, and different metric normalization methods. In both our original sample and approach and in the expanded analysis we found some metrics to be consistently predictive, such as the projected offset between the WMC and X-ray centers and the projected offset between the WMC and the MMCG/BCG, and others, such as the X-ray power ratios, the quiescent fraction, and the DS test, appear to generally have consistently limited predictive power. As discussed in Section \ref{sec:predict} these results should help to inform data-gathering strategies for classifying galaxy clusters as relaxed or disturbed in future optical/NIR or X-ray surveys. We take here each in turn. 


The case of optical/NIR surveys where accompanying X-ray data is limited is perhaps where the results of this analysis are most valuable as no information on the 
virialization state of clusters from deviations in $L_x$-$T$ space are available. In such cases, the primary predictive
metric, the projected offset between the WMC and the X-ray center is not available. However, performing enough spectroscopy to unambiguously confirm the MMCG/BCG and to estimate 
both the WMC and the systemic velocity of the cluster allows for the estimation of two metrics with higher predictive power (MMCG/BCG to WMC Dist. and MMCG/BCG Vel. Off.). From 
tests on the member populations of the ORELSE clusters, the WMC and systemic velocity can be determined at high precision ($\la50$ kpc and $\la$100 km s$^{-1}$, respectively)
from limited spectroscopy which focuses on the most massive/luminous members. However, we note that metrics which rely on a single galaxy, i.e., the MMCG/BCG, are subject to serious 
uncertainty when spectroscopy is limited, too narrow in its selection (e.g., brightness or color range, spatial coverage), or not well-informed by imaging data. Under these circumstances 
the true MMCG/BCG can easily escape spectroscopic detection resulting in potentially catastrophic consequences. If further spectroscopy is performed and if the optical/NIR imaging 
allows for the calculation of photo-$z$s to mitigate the number of bluer interlopers, such spectroscopy should focus on equally targeting blue and red photo-$z$ member galaxies. 
Adopting this strategy would allow for the calculation of the velocity center and dispersion offsets between the two populations, both of which are moderate predictors of virialization. The 
former is considerably easier to obtain as it is, at least in our cluster sample, more robust to changes in sample size and outliers and can be estimated with high precision 
with a small number of galaxies. The quiescent fraction and results from DS tests both require extensive spectroscopy to meaningfully constrain and appear to have the
least predictive power of all optical metrics. The most powerful leverage, however, comes if shallow X-ray observations are added onto the optical/NIR data. These observations need not
be deep enough to measure $L_X$ or $T_X$, but just deep enough to obtain a centroid of the diffuse ICM emission. Under this scenario, the projected distance between the WMC and 
the X-ray center, by far the most predictive metric, can be measured and used in conjunction with information on the MMCG/BCG to estimate the level of virialization with a high level of reliability. 

The case of large-scale X-ray surveys with limited or shallow accompanying optical/NIR imaging is one of limited applicability to both present and future 
cluster surveys. However, while deep, multi-band optical/NIR imaging, if it does not already exist, should become available for the vast majority of X-ray detected clusters 
over the course of the next decade, spectroscopic followup will likely not be as readily available. In such cases it is worth considering what populations are most valuable to
target spectroscopically in order to most efficiently and reliably constrain the virialization state of the cluster. While, in principle, offsets relative to
fiducial $L_x$-$T$ relations can be used for such clusters, we note that deep X-ray imaging is required to meaningfully estimate an X-ray temperature for $z\sim1$ clusters 
and, as shown in Table \ref{offsettab}, clusters which show unremarkable $L_x$-$T$ offsets
can sometimes exhibit severe offsets in the other scaling relations. The primary goal of any spectroscopic followup should, again, be aimed at unambiguously confirming the MMCG/BCG 
as well as the most massive/luminous member galaxies. With this, four of the most important metrics, the projected offsets between the MMCG/BCG, the WMCs, and the X-ray centers 
as well as the velocity offset between the MMCG/BCG and the systemic velocity can 
be calculated to a high level of precision. Again, any additional spectroscopy should focus on splitting spectroscopic targets between blue and red galaxies, with the primary
aim being to calculate the offsets between the mean velocities of the two sub-samples. It is important
to note this strategy only requires knowledge of the center of the resolved X-ray emission, which means that even shallow X-ray data along with limited, intelligently-targeted 
spectroscopy would suffice to place strong constraints on the virialization state of a cluster. While deep X-ray data can be leveraged for other purposes, e.g., to determine
the total mass of a cluster under certain assumptions, our results suggest that obtaining deeper X-ray imaging for the purposes of determining the X-ray luminosity/temperature 
of the ICM or to calculate meaningful power ratios, is, at least for the purposes of estimating the virialization state of a cluster, far too expensive for the minimal gain it 
provides.


\bigskip

\section*{Acknowledgements}

\footnotesize{
This material is based upon work supported by the National Aeronautics and Space Administration under NASA Grant Number NNX15AK92G. Part of the work presented herein is supported by the National Science Foundation under Grant No. 1411943. The authors thank Kirpal Nandra and Antonis Georgakakis for providing the Imperial reduction pipeline and their ongoing support of the software. The authors also thank the anonymous referee for suggestions which allowed us to catch several errors and spurred a vast improvement in the content and scope of the paper. Work presented here is based in part on data collected at Subaru Telescope as well as archival data obtained from the SMOKA, which is operated by the Astronomy Data Center, National Astronomical Observatory of Japan. This work is based in part on observations made with the Large Format Camera mounted on the 200-inch Hale Telescope at Palomar Observatory, owned and operated by the California Institute of Technology. A subset of observations were obtained with WIRCam, a joint project of CFHT, Taiwan, Korea, Canada, France, at the Canada-France-Hawaii Telescope (CFHT) which is operated by the National Research Council (NRC) of Canada, the Institut National des Sciences de l'Univers of the Centre National de la Recherche Scientifique of France, and the University of Hawaii. UKIRT is supported by NASA and operated under an agreement among the University of Hawaii, the University of Arizona, and Lockheed Martin Advanced Technology Center; operations are enabled through the cooperation of the East Asian Observatory. This work is based in part on observations made with the Spitzer Space Telescope, which is operated by the Jet Propulsion Laboratory, California Institute of Technology under a contract with NASA. The spectroscopic data presented herein were obtained at the W.M. Keck Observatory, which is operated as a scientific partnership among the California Institute of Technology, the University of California and the National Aeronautics and Space Administration. The Observatory was made possible by the generous financial support of the W.M. Keck Foundation. The authors wish to recognize and acknowledge the very significant cultural role and reverence that the summit of Mauna Kea has always had within the indigenous Hawaiian community.  We are most fortunate to have the opportunity to conduct observations from this mountain. }
 
\bibliographystyle{mn2e}

\bibliography{rum}

\end{document}